\documentclass[twocolumn,astrosymb]{aastex631}

\usepackage[utf8]{inputenc}
\usepackage{amsmath}

\newcommand{\Var}{\operatorname{Var}}

\shorttitle{}
\shortauthors{IXPE Collaboration et al.}
\graphicspath{{./}{Figures/}}

\begin{document}

\title{X-ray polarization detection of Cassiopeia A with IXPE}

\author{Jacco Vink}
\correspondingauthor{Jacco Vink}
\email{j.vink@uva.nl}
\affiliation{Anton Pannekoek Institute for Astronomy \& GRAPPA, University of Amsterdam, Science Park 904, 1098 XH Amsterdam, The Netherlands}
\author{Dmitry Prokhorov}
\affiliation{Anton Pannekoek Institute for Astronomy \& GRAPPA, University of Amsterdam, Science Park 904, 1098 XH Amsterdam, The Netherlands}
\author{Riccardo Ferrazzoli}
\affiliation{INAF Istituto di Astrofisica e Planetologia Spaziali, Via del Fosso del Cavaliere 100, 00133 Roma, Italy}
\author{Patrick Slane}
\affiliation{Center for Astrophysics, Harvard \& Smithsonian, 60 Garden St, Cambridge, MA 02138, USA}
\author{Ping Zhou}
\affiliation{School of Astronomy and Space Science, Nanjing University, Nanjing 210023, PR China}
\author{Kazunori Asakura}
\affiliation{Osaka University, Graduate School of Science, Osaka, Japan}
\author{Luca Baldini}
\affiliation{Istituto Nazionale di Fisica Nucleare, Sezione di Pisa, Largo B. Pontecorvo 3, 56127 Pisa, Italy}
\affiliation{Dipartimento di Fisica, Università di Pisa, Largo B. Pontecorvo 3, 56127 Pisa, Italy}
\author{Niccolò Bucciantini}
\affiliation{INAF Osservatorio Astrofisico di Arcetri, Largo Enrico Fermi 5, 50125 Firenze, Italy}
\affiliation{Dipartimento di Fisica e Astronomia, Università degli Studi di Firenze, Via Sansone 1, 50019 Sesto Fiorentino (FI), Italy}
\affiliation{Istituto Nazionale di Fisica Nucleare, Sezione di Firenze, Via Sansone 1, 50019 Sesto Fiorentino (FI), Italy}
\author{Enrico Costa}
\affiliation{INAF Istituto di Astrofisica e Planetologia Spaziali, Via del Fosso del Cavaliere 100, 00133 Roma, Italy}

\author{Alessandro Di Marco}
\affiliation{INAF Istituto di Astrofisica e Planetologia Spaziali, Via del Fosso del Cavaliere 100, 00133 Roma, Italy}

\author{Jeremy Heyl}
\affiliation{University of British Columbia, Vancouver, BC V6T 1Z4, Canada}
\author{Frédéric Marin}
\affiliation{Université de Strasbourg, CNRS, Observatoire Astronomique de Strasbourg, UMR 7550, 67000 Strasbourg, France}
\author{Tsunefumi Mizuno}
\affiliation{Hiroshima Astrophysical Science Center, Hiroshima University, 1-3-1 Kagamiyama, Higashi-Hiroshima, Hiroshima 739-8526, Japan}
\author{C.-Y. Ng}
\affiliation{Department of Physics, The University of Hong Kong, Pokfulam, Hong Kong}
\author{Melissa Pesce-Rollins}
\affiliation{Istituto Nazionale di Fisica Nucleare, Sezione di Pisa, Largo B. Pontecorvo 3, 56127 Pisa, Italy}

\author{Brian D. Ramsey}
\affiliation{NASA Marshall Space Flight Center, Huntsville, AL 35812, USA}

\author{John Rankin}
\affiliation{INAF Istituto di Astrofisica e Planetologia Spaziali, Via del Fosso del Cavaliere 100, 00133 Roma, Italy}
\author{Ajay Ratheesh}
\affiliation{INAF Istituto di Astrofisica e Planetologia Spaziali, Via del Fosso del Cavaliere 100, 00133 Roma, Italy}

\author{Carmelo Sgrò}
\affiliation{Istituto Nazionale di Fisica Nucleare, Sezione di Pisa, Largo B. Pontecorvo 3, 56127 Pisa, Italy}
\author{Paolo Soffitta}
\affiliation{INAF Istituto di Astrofisica e Planetologia Spaziali, Via del Fosso del Cavaliere 100, 00133 Roma, Italy}
\author{Douglas A. Swartz}
\affiliation{NASA Marshall Space Flight Center, Huntsville, AL 35812, USA}
\author{Toru Tamagawa}
\affiliation{RIKEN Cluster for Pioneering Research, 2-1 Hirosawa, Wako, Saitama 351-0198, Japan}
\author{Martin C. Weisskopf}
\affiliation{NASA Marshall Space Flight Center, Huntsville, AL 35812, USA}
\author{Yi-Jung Yang}
\affiliation{Department of Physics, The University of Hong Kong, Pokfulam, Hong Kong}
\affiliation{Laboratory for Space Research, The University of Hong Kong, Hong Kong}
\author{Ronaldo Bellazzini}
\affiliation{Istituto Nazionale di Fisica Nucleare, Sezione di Pisa, Largo B. Pontecorvo 3, 56127 Pisa, Italy}
\author{Raffaella Bonino}
\affiliation{Istituto Nazionale di Fisica Nucleare, Sezione di Torino, Via Pietro Giuria 1, 10125 Torino, Italy}
\affiliation{Dipartimento di Fisica, Università degli Studi di Torino, Via Pietro Giuria 1, 10125 Torino, Italy}
\author{Elisabetta Cavazzuti}
\affiliation{ASI - Agenzia Spaziale Italiana, Via del Politecnico snc, 00133 Roma, Italy}
\author{Luigi Costamante}
\affiliation{ASI - Agenzia Spaziale Italiana, Via del Politecnico snc, 00133 Roma, Italy}
\author{Niccolò Di Lalla}
\affiliation{Department of Physics and Kavli Institute for Particle Astrophysics and Cosmology, Stanford University, Stanford, California 94305, USA}
\author{Luca Latronico}
\affiliation{Istituto Nazionale di Fisica Nucleare, Sezione di Torino, Via Pietro Giuria 1, 10125 Torino, Italy}
\author{Simone Maldera}
\affiliation{Istituto Nazionale di Fisica Nucleare, Sezione di Torino, Via Pietro Giuria 1, 10125 Torino, Italy}
\author{Alberto Manfreda}
\affiliation{Istituto Nazionale di Fisica Nucleare, Sezione di Pisa, Largo B. Pontecorvo 3, 56127 Pisa, Italy}
\author{Francesco Massaro}
\affiliation{Istituto Nazionale di Fisica Nucleare, Sezione di Torino, Via Pietro Giuria 1, 10125 Torino, Italy}
\affiliation{Dipartimento di Fisica, Università degli Studi di Torino, Via Pietro Giuria 1, 10125 Torino, Italy}
\author{Ikuyuki Mitsuishi}
\affiliation{Graduate School of Science, Division of Particle and Astrophysical Science, Nagoya University, Furo-cho, Chikusa-ku, Nagoya, Aichi 464-8602, Japan}
\author{Nicola Omodei}
\affiliation{Department of Physics and Kavli Institute for Particle Astrophysics and Cosmology, Stanford University, Stanford, California 94305, USA}
\author{Chiara Oppedisano}
\affiliation{Istituto Nazionale di Fisica Nucleare, Sezione di Torino, Via Pietro Giuria 1, 10125 Torino, Italy}

\author{Silvia Zane}
\affiliation{Mullard Space Science Laboratory, University College London, Holmbury St Mary, Dorking, Surrey RH5 6NT, UK}
\author{Ivan Agudo}
\affiliation{Instituto de Astrofísicade Andalucía, IAA-CSIC, Glorieta de la Astronomía s/n, 18008 Granada, Spain}
\author{Lucio A. Antonelli}
\affiliation{INAF Osservatorio Astronomico di Roma, Via Frascati 33, 00078 Monte Porzio Catone (RM), Italy}
\affiliation{Space Science Data Center, Agenzia Spaziale Italiana, Via del Politecnico snc, 00133 Roma, Italy}
\author{Matteo Bachetti}
\affiliation{INAF Osservatorio Astronomico di Cagliari, Via della Scienza 5, 09047 Selargius (CA), Italy}
\author{Wayne H. Baumgartner}
\affiliation{NASA Marshall Space Flight Center, Huntsville, AL 35812, USA}
\author{Stefano Bianchi}
\affiliation{Dipartimento di Matematica e Fisica, Universit\`a degli Studi Roma Tre, Via della Vasca Navale 84, 00146 Roma, Italy}
\author{Stephen D. Bongiorno}
\affiliation{NASA Marshall Space Flight Center, Huntsville, AL 35812, USA}
\author{Alessandro Brez}
\affiliation{Istituto Nazionale di Fisica Nucleare, Sezione di Pisa, Largo B. Pontecorvo 3, 56127 Pisa, Italy}
\author{Fiamma Capitanio}
\affiliation{INAF Istituto di Astrofisica e Planetologia Spaziali, Via del Fosso del Cavaliere 100, 00133 Roma, Italy}
\author{Simone Castellano}
\affiliation{Istituto Nazionale di Fisica Nucleare, Sezione di Pisa, Largo B. Pontecorvo 3, 56127 Pisa, Italy}
\author{Stefano Ciprini}
\affiliation{Istituto Nazionale di Fisica Nucleare, Sezione di Roma "Tor Vergata", Via della Ricerca Scientifica 1, 00133 Roma, Italy}
\affiliation{Space Science Data Center, Agenzia Spaziale Italiana, Via del Politecnico snc, 00133 Roma, Italy}
\author{Alessandra De Rosa}
\affiliation{INAF Istituto di Astrofisica e Planetologia Spaziali, Via del Fosso del Cavaliere 100, 00133 Roma, Italy}
\author{Ettore Del Monte}
\affiliation{INAF Istituto di Astrofisica e Planetologia Spaziali, Via del Fosso del Cavaliere 100, 00133 Roma, Italy}
\author{Laura Di Gesu}
\affiliation{ASI - Agenzia Spaziale Italiana, Via del Politecnico snc, 00133 Roma, Italy}

\author{Immacolata Donnarumma}
\affiliation{ASI - Agenzia Spaziale Italiana, Via del Politecnico snc, 00133 Roma, Italy}
\author{Victor Doroshenko}
\affiliation{Institut f\"ur Astronomie und Astrophysik, Universität Tübingen, Sand 1, 72076 T\"ubingen, Germany}
\affiliation{Space Research Institute of the Russian Academy of Sciences, Profsoyuznaya Str. 84/32, Moscow 117997, Russia}
\author{Michal Dovčiak}
\affiliation{Astronomical Institute of the Czech Academy of Sciences, Boční II 1401/1, 14100 Praha 4, Czech Republic}
\author{Steven R. Ehlert}
\affiliation{NASA Marshall Space Flight Center, Huntsville, AL 35812, USA}
\author{Teruaki Enoto}
\affiliation{RIKEN Cluster for Pioneering Research, 2-1 Hirosawa, Wako, Saitama 351-0198, Japan}
\author{Yuri Evangelista}
\affiliation{INAF Istituto di Astrofisica e Planetologia Spaziali, Via del Fosso del Cavaliere 100, 00133 Roma, Italy}
\author{Sergio Fabiani}
\affiliation{INAF Istituto di Astrofisica e Planetologia Spaziali, Via del Fosso del Cavaliere 100, 00133 Roma, Italy}
\author{Javier A. Garcia}
\affiliation{California Institute of Technology, Pasadena, CA 91125, USA}
\author{Shuichi Gunji}
\affiliation{Yamagata University,1-4-12 Kojirakawa-machi, Yamagata-shi 990-8560, Japan}
\author{Kiyoshi Hayashida}
\affiliation{Osaka University, 1-1 Yamadaoka, Suita, Osaka 565-0871, Japan}
\author{Wataru Iwakiri}
\affiliation{Department of Physics, Faculty of Science and Engineering, Chuo University, 1-13-27 Kasuga, Bunkyo-ku, Tokyo 112-8551, Japan}
\author{Svetlana G. Jorstad}
\affiliation{Institute for Astrophysical Research, Boston University, 725 Commonwealth Avenue, Boston, MA 02215, USA}
\affiliation{Department of Astrophysics, St. Petersburg State University, Universitetsky pr. 28, Petrodvoretz, 198504 St. Petersburg, Russia}
\author{Vladimir Karas}
\affiliation{Astronomical Institute of the Czech Academy of Sciences, Boční II 1401/1, 14100 Praha 4, Czech Republic}
\author{Takao Kitaguchi}
\affiliation{RIKEN Cluster for Pioneering Research, 2-1 Hirosawa, Wako, Saitama 351-0198, Japan}
\author{Jeffery J. Kolodziejczak}
\affiliation{NASA Marshall Space Flight Center, Huntsville, AL 35812, USA}
\author{Henric Krawczynski}
\affiliation{Physics Department and McDonnell Center for the Space Sciences, Washington University in St. Louis, St. Louis, MO 63130, USA}
\author{Fabio La Monaca}
\affiliation{INAF Istituto di Astrofisica e Planetologia Spaziali, Via del Fosso del Cavaliere 100, 00133 Roma, Italy}
\author{Ioannis Liodakis}
\affiliation{Finnish Centre for Astronomy with ESO,  20014 University of Turku, Finland}
\author{Andrea Marinucci}
\affiliation{ASI - Agenzia Spaziale Italiana, Via del Politecnico snc, 00133 Roma, Italy}
\author{Alan P. Marscher}
\affiliation{Institute for Astrophysical Research, Boston University, 725 Commonwealth Avenue, Boston, MA 02215, USA}
\author{Herman L. Marshall}
\affiliation{MIT Kavli Institute for Astrophysics and Space Research, Massachusetts Institute of Technology, 77 Massachusetts Avenue, Cambridge, MA 02139, USA}
\author{Giorgio Matt}
\affiliation{Dipartimento di Matematica e Fisica, Universit\`a degli Studi Roma Tre, Via della Vasca Navale 84, 00146 Roma, Italy}
\author{Fabio Muleri}
\affiliation{INAF Istituto di Astrofisica e Planetologia Spaziali, Via del Fosso del Cavaliere 100, 00133 Roma, Italy}
\author{Stephen L. O'Dell}
\affiliation{NASA Marshall Space Flight Center, Huntsville, AL 35812, USA}
\author{Alessandro Papitto}
\affiliation{INAF Osservatorio Astronomico di Roma, Via Frascati 33, 00078 Monte Porzio Catone (RM), Italy}
\author{George G. Pavlov}
\affiliation{Department of Astronomy and Astrophysics, Pennsylvania State University, University Park, PA 16802, USA}
\author{Abel L. Peirson}
\affiliation{Department of Physics and Kavli Institute for Particle Astrophysics and Cosmology, Stanford University, Stanford, California 94305, USA}
\author{Matteo Perri}
\affiliation{Space Science Data Center, Agenzia Spaziale Italiana, Via del Politecnico snc, 00133 Roma, Italy}
\affiliation{INAF Osservatorio Astronomico di Roma, Via Frascati 33, 00078 Monte Porzio Catone (RM), Italy}
\author{Maura Pilia}
\affiliation{INAF Osservatorio Astronomico di Cagliari, Via della Scienza 5, 09047 Selargius (CA), Italy}
\author{Andrea Possenti}
\affiliation{INAF Osservatorio Astronomico di Cagliari, Via della Scienza 5, 09047 Selargius (CA), Italy}
\author{Juri Poutanen}
\affiliation{Department of Physics and Astronomy, 20014 University of Turku, Finland}
\affiliation{Space Research Institute of the Russian Academy of Sciences, Profsoyuznaya Str. 84/32, Moscow 117997, Russia}
\author{Simonetta Puccetti}
\affiliation{Space Science Data Center, Agenzia Spaziale Italiana, Via del Politecnico snc, 00133 Roma, Italy}

\author{Roger W. Romani}
\affiliation{Department of Physics and Kavli Institute for Particle Astrophysics and Cosmology, Stanford University, Stanford, California 94305, USA}
\author{Gloria Spandre}
\affiliation{Istituto Nazionale di Fisica Nucleare, Sezione di Pisa, Largo B. Pontecorvo 3, 56127 Pisa, Italy}
\author{Fabrizio Tavecchio}
\affiliation{INAF Osservatorio Astronomico di Brera, Via E. Bianchi 46, 23807 Merate (LC), Italy}
\author{Roberto Taverna}
\affiliation{Dipartimento di Fisica e Astronomia, Università degli Studi di Padova, Via Marzolo 8, 35131 Padova, Italy}
\author{Yuzuru Tawara}
\affiliation{Graduate School of Science, Division of Particle and Astrophysical Science, Nagoya University, Furo-cho, Chikusa-ku, Nagoya, Aichi 464-8602, Japan}
\author{Allyn F. Tennant}
\affiliation{NASA Marshall Space Flight Center, Huntsville, AL 35812, USA}
\author{Nicolas E. Thomas}
\affiliation{NASA Marshall Space Flight Center, Huntsville, AL 35812, USA}
\author{Francesco Tombesi}
\affiliation{Dipartimento di Fisica, Universit\`a degli Studi di Roma "Tor Vergata", Via della Ricerca Scientifica 1, 00133 Roma, Italy}
\affiliation{Istituto Nazionale di Fisica Nucleare, Sezione di Roma "Tor Vergata", Via della Ricerca Scientifica 1, 00133 Roma, Italy}
\affiliation{Department of Astronomy, University of Maryland, College Park, Maryland 20742, USA}
\author{Alessio Trois}
\affiliation{INAF Osservatorio Astronomico di Cagliari, Via della Scienza 5, 09047 Selargius (CA), Italy}
\author{Sergey Tsygankov}
\affiliation{Department of Physics and Astronomy, 20014 University of Turku, Finland}
\affiliation{Space Research Institute of the Russian Academy of Sciences, Profsoyuznaya Str. 84/32, Moscow 117997, Russia}
\author{Roberto Turolla}
\affiliation{Dipartimento di Fisica e Astronomia, Università degli Studi di Padova, Via Marzolo 8, 35131 Padova, Italy}
\affiliation{Mullard Space Science Laboratory, University College London, Holmbury St Mary, Dorking, Surrey RH5 6NT, UK}
\author{Kinwah Wu}
\affiliation{Mullard Space Science Laboratory, University College London, Holmbury St Mary, Dorking, Surrey RH5 6NT, UK}
\author{Fei Xie}
\affiliation{Guangxi Key Laboratory for Relativistic Astrophysics, School of Physical Science and Technology, Guangxi University, Nanning 530004, China}

\begin{abstract}
We report  on a  $\sim 5\sigma$ detection of  polarized 3--6~keV X-ray emission from the  supernova remnant  Cassiopeia  A with
the Imaging X-ray Polarimetry Explorer (IXPE). 
The overall polarization degree of  $1.8 \pm 0.3$\% is detected by summing over a large region, assuming circular symmetry for the polarization vectors.
The measurements imply an average polarization degree for the synchrotron component of $\sim 2.5$\%, and close to 5\% for the X-ray synchrotron-domimated forward-shock region.
These numbers are  based on an assessment of the thermal and non-thermal radiation contributions, for which we used a detailed spatial-spectral model based on Chandra X-ray data.
A pixel-by-pixel search for polarization provides a few  tentative detections from discrete regions at the $\sim 3\sigma$ confidence level. 
Given the number of pixels, the significance is  {insufficient} to claim a detection for individual pixels, but 
 implies considerable turbulence on scales smaller than the angular resolution. 
Cas A's  X-ray continuum emission is dominated by
synchrotron radiation from regions  within $\lesssim 10^{17}$~cm of the forward- and  reverse shocks. 
We find that i) the measured polarization angle corresponds to a radially-oriented magnetic field, similar to what has been inferred from radio observations; 
ii) the X-ray polarization degree is  lower than in the radio band ($\sim 5$\%).
Since shock compression should impose a tangential magnetic field structure, the IXPE results imply that 
magnetic-fields are reoriented within $\sim 10^{17}$~cm  of the shock. If the magnetic-field alignment is due to locally enhanced acceleration near
quasi-parallel shocks, the preferred X-ray polarization angle suggests a size of $3\times 10^{16}$~cm for cells with radial magnetic fields.
\end{abstract}



\section{Introduction} \label{sec:intro}
Supernova remnants (SNRs) have long been known to be sources of radio synchrotron radiation \citep{shklovsky54}, 
emitted by relativistic electrons, {which are} now recognized
to be accelerated through the diffusive shock acceleration (DSA) process \citep[e.g.,][]{malkov01}. {According
to DSA {theory}, energetic, charged particles  move diffusively through the plasma---due to scattering on magnetic-field fluctuations---and gain energy by repeatedly crossing the shock front, with each
shock crossing providing  a few percent gain in momentum.}
The presence of relativistic electrons has been key for identifying SNR shocks---as opposed to the supernova explosions themselves--- as  locations for  Galactic cosmic-ray acceleration.
Although the link between cosmic rays and SNRs has also been confirmed by gamma-ray detections of many SNRs \citep[e.g.][for reviews]{helder12b,funk17},
important information about the cosmic-ray acceleration process and shock conditions has been obtained from the detection and characterisation
of X-ray synchrotron emission from young SNRs, starting with the discovery of X-ray synchrotron radiation from the rims of SN\,1006 \citep{koyama95}.

The $\sim$10--100~TeV electrons responsible for the X-ray synchrotron radiation have radiative-loss timescales that are short compared to the SNR ages:  $\tau_{\rm loss}\approx 12.5 E_{14}^{-1}B_{-4}^{-2}$~yr,
with $E_{14}$ the  electron energy in units of 100~TeV and $B_{-4}\equiv B/10^{-4}~{\rm G}$. X-ray synchrotron radiation from SNRs, therefore, indicates that
the 10--100~TeV electrons {have been accelerated recently} and on short time scales. According to DSA theory, {reaching such high energies} requires highly turbulent fields --- $|\delta B/B|\sim 1$ --- resulting in 
 mean-free-path lengths of the order of the gyroradius for these electrons.

Synchrotron radiation is intrinsically polarized at the $\sim 70$\% level \citep{ginzburg65}, with the polarization angle informing us about the local magnetic-field orientation, and the polarization
degree being dependent on the uniformity of the magnetic-field orientation along the line of sight of observed regions. 
Mature SNRs  {($\gtrsim 2000$~yr)} tend to have magnetic fields that are tangentially oriented with respect to the radial direction, 
whereas young SNRs have radially oriented magnetic fields \citep{dickel76}. See also the review by \citet{dubner15}.
The {polarization degree in the radio band} is also different between young and mature SNRs, with mature SNRs generally having  polarization degrees above  $10$\%,  
while young SNRs {($\lesssim 2000$~yr)} 
generally have polarization degrees below 10\% {\citep{dickel90,sun11}}.
The radial orientation of the magnetic fields in young SNRs
is not well understood from a theoretical point of view, but see \citet{zirakashvili08a,inoue13,west17} for a number of hypotheses. It is not known whether the radially oriented field is immediately
established at the shock front. It may well be that the fields closer to the shock fronts are tangentially oriented \citep{jun96,bykov20}, 
since the shock compresses, and thus enhances, the tangential component of the pre-shock magnetic field.

An {object} that has in many ways been central to the discussion of radio and X-ray synchrotron radiation from SNRs is the young ($\sim 350$~yr) and bright core-collapse SNR
Cassiopeia A (Cas A), located at a distance of 3.4~kpc \citep{reed95}.  
Cas A has a radially oriented magnetic-field \citep[e.g.][]{rosenberg70b,braun87b},  and a polarization degree in the radio band of only 5\% in the bright shell, and 8--10\% in the outer regions
 \citep[e.g.][]{anderson95b}. 
{The   $\sim 5$\% level of polarization degree has  been established at radio frequencies from 20--100 GHz \citep{mayer68,flett79,kenny85} and even in the infrared, at 2.2~${\rm \mu m}$, for a segment in the northwest of Cas A \citep{jones03}.  At these high frequencies Faraday rotation is negligible---see also \citep{kenny85}---and the only factor contributing to the low polarization degree is the non-uniformity of the magnetic-field directions within a angular resolution element and integrated along the line of sight.}
 
X-ray polarization measurements are in that respect highly interesting.
First of all, 
the rapid radiative losses of the 10--100 TeV electrons mean  that X-ray synchrotron emission originates from plasma that is confined to  thin regions downstream of the shock, with a typical width of $l_{\rm loss}\approx \Delta v\tau_{\rm loss}$, with $\Delta v=\frac{1}{4}V_{\rm sh}$. For Cas A the shock velocity is $V_{\rm sh}\approx 5000$--$6000$~km\,s$^{-1}$ \citep[e.g][]{patnaude09, vink22}.
Indeed, the observed widths {of the X-ray synchrotron filaments of} 1\arcsec--2\arcsec\ ($l\approx 10^{17}$~cm) suggests that the local, downstream magnetic-field strengths are  $B\approx 250$--$550~{\rm \mu G}$ \citep{vink03a,bamba05,voelk05,ballet06,helder12b}.
As a result, X-ray synchrotron emission originates from a much smaller volume than the radio synchrotron emission, which originates from the entire shell with lines of sight of $l\gtrsim 10^{18}$~cm.
The shorter pathlengths probed in X-ray synchrotron radiation should
result in less depolarization caused by variations in magnetic-field orientations along the line
of sight. 
{In addition,}
due to the steepness of the X-ray synchrotron spectrum, the intrinsic, maximum polarization degree could be larger than in the radio \citep{ginzburg65,bykov09}.

However, 
as already stated above, the electron energies of  10--100~TeV necessary  for X-ray synchrotron radiation require DSA in the presence of very turbulent magnetic fields immediately upstream
of the shock fronts. 
Although  shock  compression enhances the tangential components of the magnetic field, a high level of magnetic-field turbulence may persist in the downstream region, perhaps even isotropizing the magnetic field
due to non-linear interactions of the fluctuations downstream of the shock \citep{bykov20}. But it is also possible that the turbulence rapidly decays within the region producing the X-ray synchrotron radiation \citep{pohl05}.
{Whatever} the mechanism  that created the large scale radially oriented magnetic field inferred from radio emission {is}, this mechanism may already start to
realign magnetic fields in a radial direction immediately downstream of the shocks \citep{jun96,inoue13}.
Note that an interesting feature of the X-ray synchrotron emission from Cas A is that it does not only arise from the forward shock, but also from the reverse shock region,
in particular in the western part of the SNR \citep{helder08,uchiyama08,grefenstette15}, where the reverse shock appears to be moving toward the center instead of outward \citep{sato18,vink22}. 

With the recent launch of the Imaging X-ray Polarimetry Explorer (IXPE) \citep{weisskopf21}, we can finally measure  and map polarization with spatial resolution of a few tens of arcseconds.  
Cas A was the first science target of IXPE, and we report here the first detection of X-ray polarization from a shell-type SNR, albeit with a {surprisingly}
low polarization degree of {$\lesssim 4$\%}.
{Although IXPE cannot resolve the X-ray synchrotron filaments,}
the fact that X-ray synchrotron radiation originates from within $\sim 10^{17}$~cm downstream of the shock (i.e. in the radial direction), 
allows us  to probe the magnetic-field orientation and isotropy 
to within this distance of the shocks.

In Section~\ref{sec:data}, we describe the measurements and analysis approach and {the resulting polarization measurements.} In Section~\ref{sec:discussion} we discuss the implications of our results in the context of models for producing polarized emission from energetic particles in SNRs. Our conclusions are presented in Section~\ref{sec:conclusions}. Details on the definition and treatment of the Stokes parameters, and on treatment of the unpolarized thermal emission in Cas A are given in Appendices.

\section{Observations, data analysis and results}
\label{sec:data}

IXPE is a {NASA/ASI}\footnote{Agenzia Spaziale Italiana} {small explorer (SMEX)} mission that was launched on  December 9, 2021. The polarization sensitive X-ray detectors are gas-pixel detectors (GPD) filled with dimethyl ether  \citep{costa01,baldini21},
{which are placed at the focus of three Wolter-1 mirror module assemblies giving telescopes with angular resolutions of 24\arcsec, 29\arcsec\ and 30\arcsec\ (half-power diameters) and each with a field of view of 12.9\arcmin\ square}
{\citep[see][]{weisskopf21,soffitta21}.} 
{An X-ray} photon interacting with the gas in the GPD results in the ejection of a photoelectron {in the direction $\phi$}, 
{which is distributed as}  $\cos^2 \theta$, {where $\theta$ is}
 the polarization direction of the electromagnetic wave. {
 The photoelectron ionizes the gas and produces secondary electrons, which form a charge cloud. These electrons, after drifting and mulitplication, are then detected and 
 recorded by a finely pixellated detector plane.}
A moments analysis of the charge-cloud shape is used to reconstruct the initial
photo-electron's direction, $\phi$ {\citep[see][for details]{bellazzini03}.}

The IXPE detectors are sensitive for polarization in the X-ray range from 2--8 keV, with a mirror effective area of 590~cm$^2$ at 4.5~keV for the three telescopes combined, reduced
{to a combined 26~cm$^2$ after accounting for detector quantum efficiency.}
The polarization sensitivity improves with energy, having a  modulation factor of $\mu \sim 15$\% at 2~keV to {$\mu =$50--60\%} at 8 keV. 
The modulation factor is the amplitude of the modulation due to polarization for a 100\% polarized source in the absence of background. An ideal polarization detector will have $\mu = 100\%$.
The detection of polarization also depends on the number of detected photons, which {in turn} depends on the source spectrum, the detector efficiency and the telescope effective area; for IXPE, the optimum energy
for detecting polarization is roughly $\sim 3$~keV. Cas A's 2--8 keV spectrum is dominated by  K-shell line emission from Si, S, Ar, Ca, and Fe, and continuum emission that is {in many regions} dominated by
synchrotron radiation \citep[e.g.,][see also Appendix~\ref{app:chandra}]{helder08}. The 4--6 keV band has only weak line emission, but 
{while using}
IXPE simulations based on imaging spectroscopy with the Chandra {X-ray Observatory (Chandra for short)} (Appendix~\ref{app:chandra}), we found that a priori
the 3--6 keV band offered a better sensitivity for detecting polarization, despite the presence of Ar-K and Ca-K line emission.

\begin{figure*}
  \centerline{
     \includegraphics[trim=70 80 290 264,clip=true,height=0.29\textwidth]{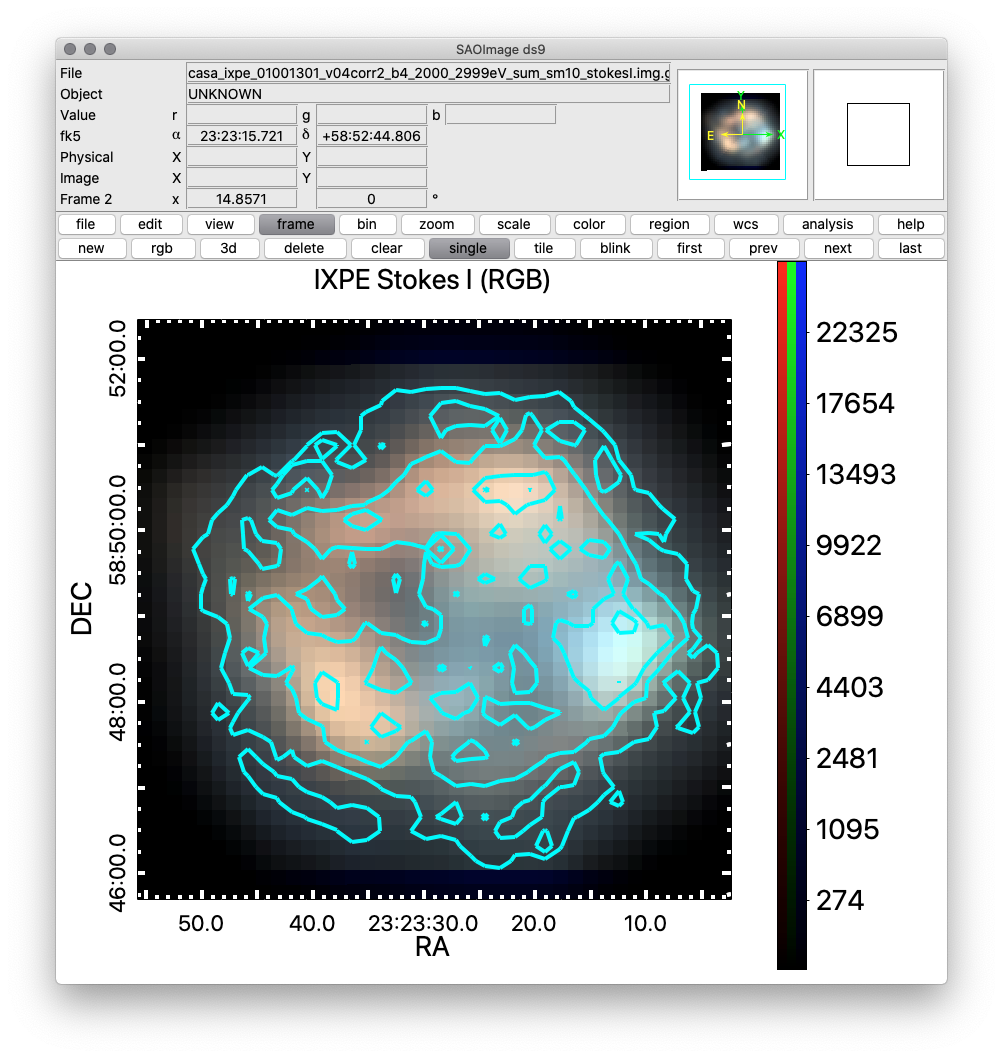}
        \includegraphics[trim=85 80 145 264,clip=true,height=0.29\textwidth]{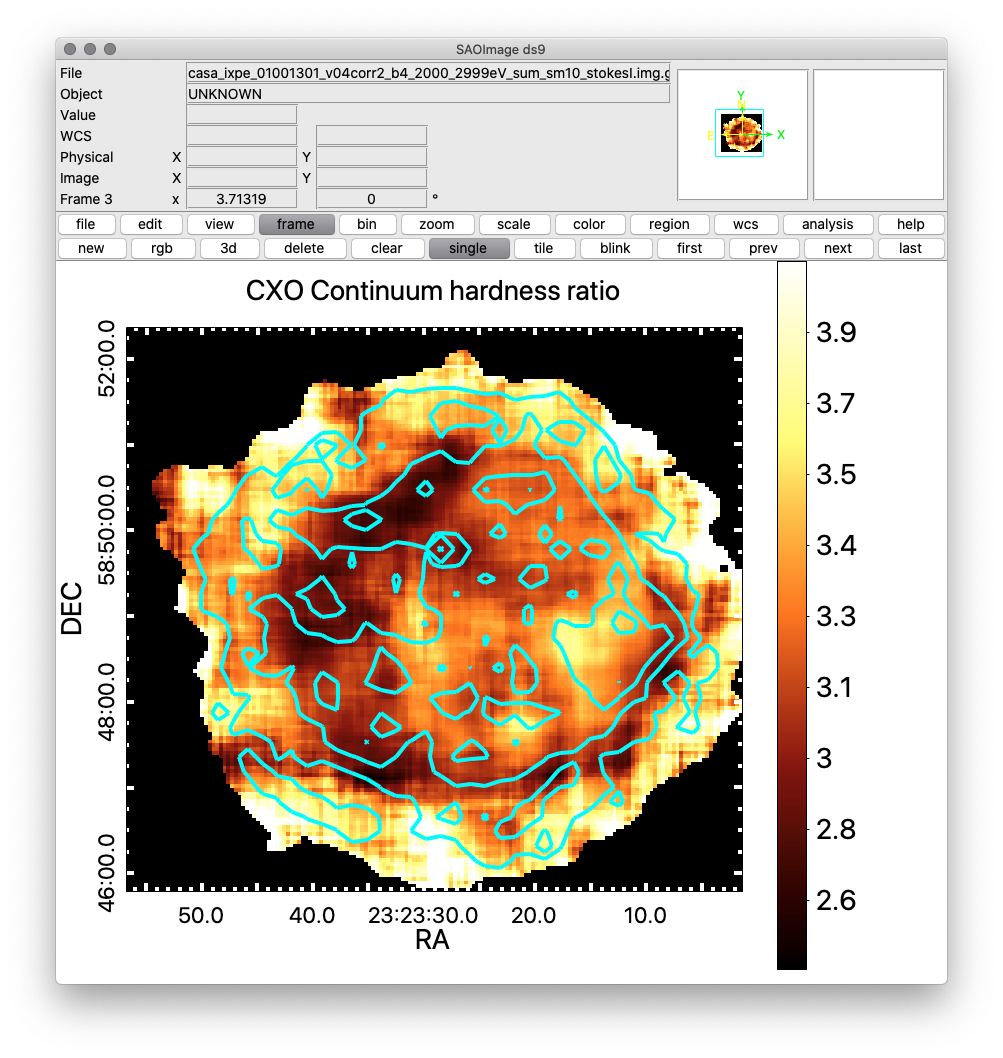}
        \includegraphics[trim=99 80 130 264,clip=true,height=0.29\textwidth]{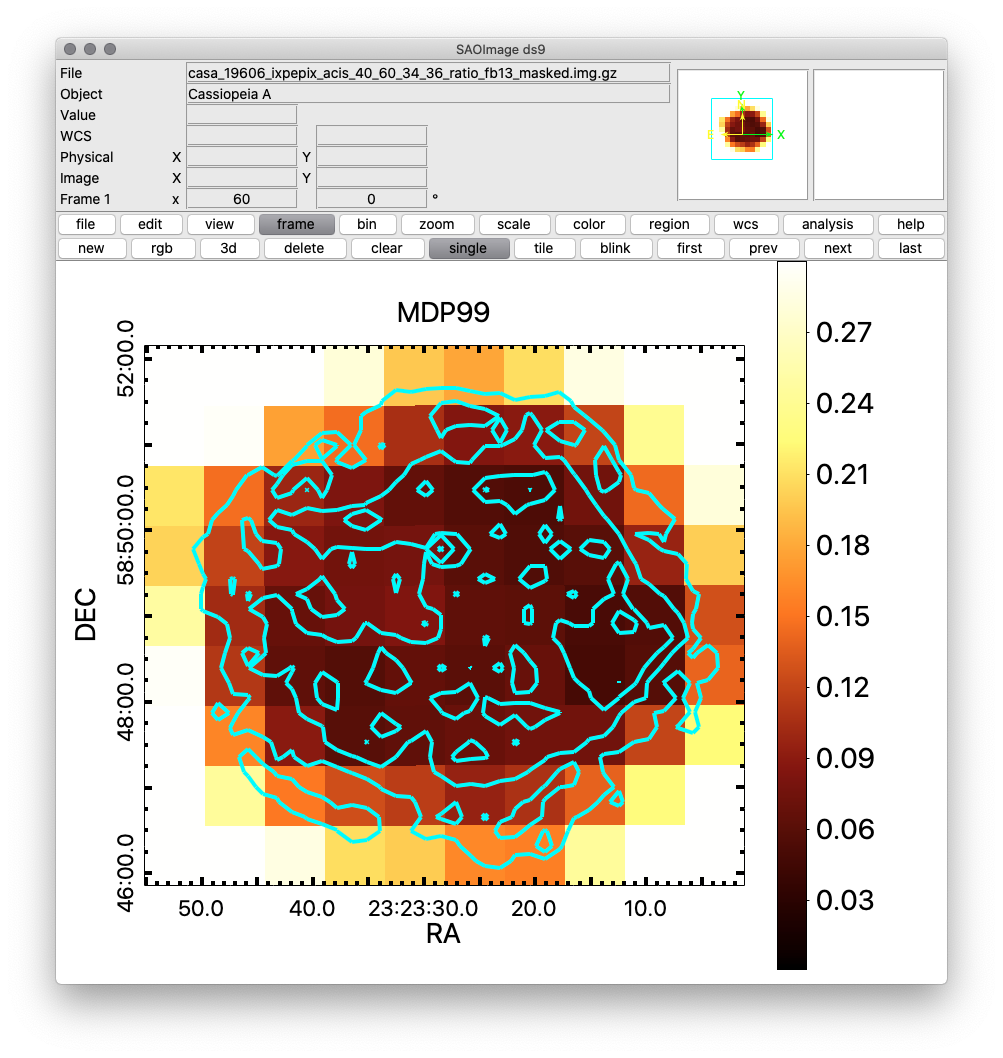}
    }
  \caption{
    \label{fig:stokesI}
    Left: IXPE three color Stokes $I$ image with square-root brightness scaling,
    based on the  2--3 keV, 3--4 keV, and
    4--6 keV bands, {combined from the three detectors}. The pixel size is 10.4\arcsec, oversampling the IXPE angular resolution by a factor $\sim 2$, and the images
    have been smoothed with a gaussian kernel with $\sigma=10.4$\arcsec. 
    Center: Hardness ratio map based on Chandra X-ray data (ObsID 19606) using the 4.0--6.0 keV map divided by the 3.4--3.6 keV map, both dominated by
    continuum emission. The harder ratios ($\gtrsim 3$) likely correspond
    to synchrotron dominated continuum emission. Fluctuations due to differential absorption across the SNR are at the 9\% level.
    Right: The MDP99 levels for      the 3--6 keV band for the
    IXPE observations of Cas A. Here the map is binned to a pixel size of 42\arcsec.
    In all panels we show the contours  based on a recent 4--6 keV Chandra map using square-root brightness scaling (ObSID 19606).
  }
\end{figure*}

Cas A was the first science target of IXPE, after first observing the calibration source {SMC X-1} {during the one month commissioning phase}. IXPE observed Cas A from January 11 to January 29, 2022, for a total time around  900~ks. 
We analyzed  the data with the
software package \texttt{ixpeobssim} \citep{baldini22},
which is used both for 
extraction of Stokes maps and spectra, as well as for 
Monte Carlo simulations.
The analysis was based on 
{
processed high-level event list (level 2) 
}
in FITS format \citep[see][]{rankin22},
which, apart from the usual columns for time, energy channel and sky and detector coordinates,
also contains columns with $q_k\equiv 2\cos 2\phi_k$ and $u_k\equiv 2\sin 2\phi_k$, with $\phi_k$ the reconstructed photo-electron direction of event number $k$. 
This definition is different from \citet{kislat15} as  the factor 2 is here part of the definition of $q_k$ and $u_k$.
Note that the values for $q_k$ and $u_k$ also contain a small correction, based on the method outlined by \citet{rankin22}, 
which is necessary to remove the effects of a  { small modulation measured for an  unpolarized source, as
found and calibrated on ground.  
The stability of these corrections have been assessed on ground. For in-flight data the stability of these {corrections has not yet been assessed}, but in the  3 to 6 keV band these effects are  minor. We can therefore state that the results of our analysis are not affected by spurious modulation effects. }
For the present analysis, the values of $q_k$ and $u_k$ are based on the moment analysis of the charge clouds. In a future data release the determination of $\phi_k$ may be improved, once
machine learning algorithms, currently  under development \citep{peirson21},  are employed.

During the {early} observations a number of calibration issues emerged and {several corrections were applied}. 
{First of all, we filtered out likely particle background events using the electron tracks, as described by \citet{xie21} (see also Di Marco et al. in preparation).
The fraction of the particle background removed was $\sim30$\%, while less than 1\% of the source was rejected.  
{We found, however, the impact of the particle background on the polarization signal of Cas A to be negligible.}
Secondly, the reconstructed energies  of the IXPE detected events in the level-2 data  
are
known to have detector-dependent and  time-dependent deviations,
due to various effects, including
charge built-up in the detectors.
Throughout the whole observation the three detectors were  calibrated during Earth occultations, by using onboard calibration sources, producing line emission at 1.7 keV (Si K$\alpha$) and 5.9 keV ($^{55}$Fe decaying into $^{55}$Mn) \citep{ferrazzoli20}. Given the linearity of the response, the two energies allow for measuring both the slope and bias of the calibration.
We reconstructed the time dependency of the channel-to-energy conversion. Thirdly, the boom connecting the X-ray telescopes with the spacecraft was affected by phase-dependent heating during its orbit around Earth.
As a result the boom experienced time modulated bending, resulting in the focal point changing in detector coordinates.
Corrections to these phase-dependent effects, some of which were also applied to the publicly released level 2 event list, 
still left an overall pointing reconstruction offset of the order of 2.5\arcmin . 
In order to remove the offset 
we employed a spatial correlation code developed
for measuring the expansion of Cas A  \citep{vink22}, and used it to register the pointing solution of each detector unit to the 2019 (ObsID 19606)  observation of Cas A with Chandra, for which we smoothed the Chandra
image with  $\sigma=10.4$\arcsec. For this we used the 4-6 keV continuum band, as the morphology of Cas A in this band is relatively independent of the energy response of the detectors.
This resulted in a pointing solution with an accuracy of about 1--2\arcsec. 
Finally, during the orbit, the pointing is affected by  a switch between  the
two different star-trackers, one placed near the {mirror module assemblies} and viewing forward, and one placed on the spacecraft and looking in opposite
direction. 
Currently this results in small periods of time with offsets in the pointing solutions.
For the present analysis we filtered these short intervals out, leaving 819~ks of effective exposure time. }

\subsection{Spatial exploration of polarization signals}
\label{sec:pixel_analysis}

Maps of the Stokes parameters $I$, $Q$, and $U$ are made with sums of $u_k$ and $q_k$ and a correction to account for the energy dependent modulation factor.
See Appendix~\ref{app:maps} for details. All maps shown are based on the summed maps of the three detector units. For the present analysis we did not use weights as we found that weighting with
the modulation factor $w_k\propto \mu_k^{-2}$ \citep{vink18a} resulted only in minor to no improvements in the statistical significance.

\begin{figure}
\includegraphics[trim=0 0 0 0,clip=true,width=\columnwidth]{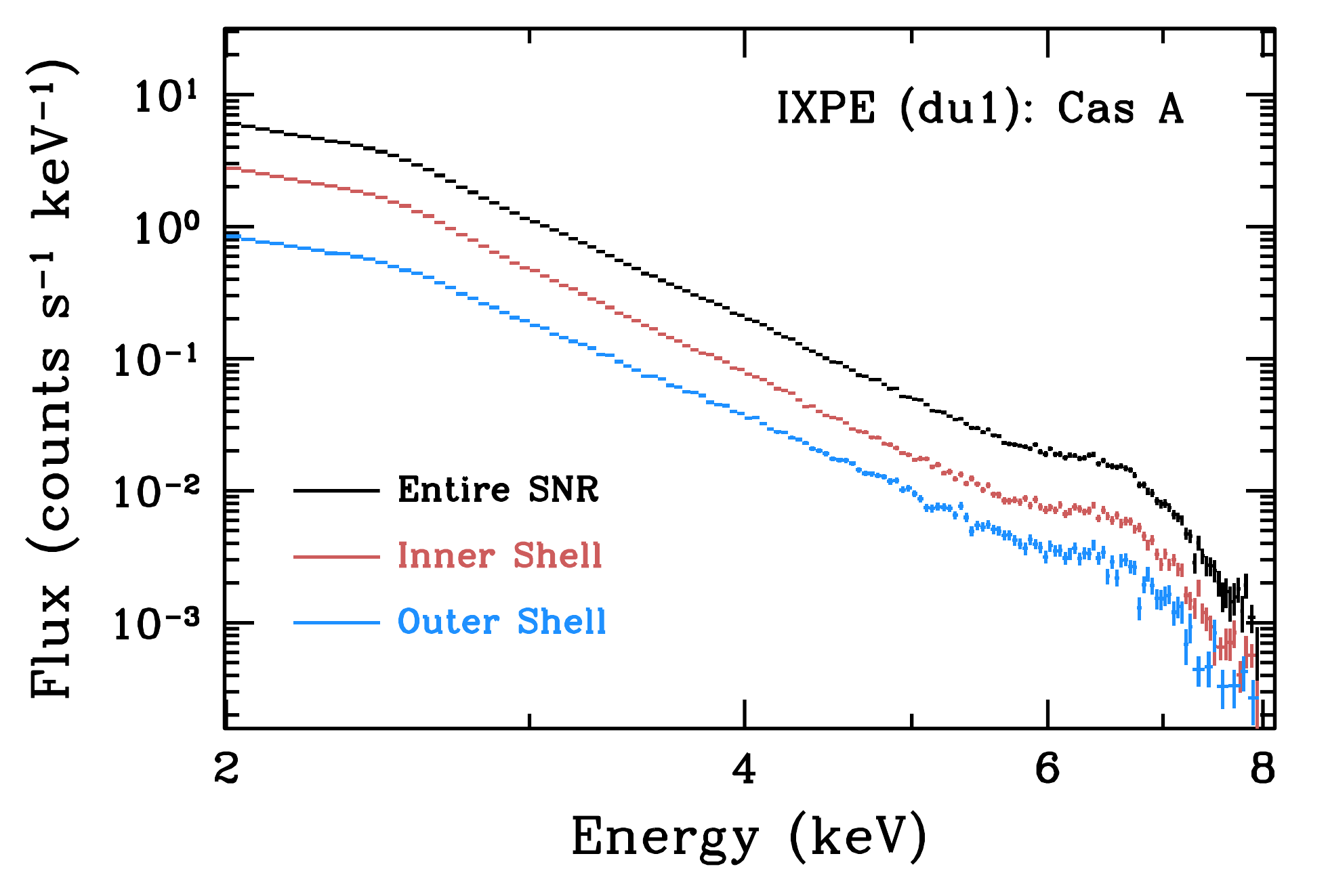}
\caption{\label{fig:spectra}
  X-ray spectra of Cas A as detected by detector unit 1 of IXPE for three
  different extraction regions. The spectrum oversamples the spectral resolution, which is $\Delta E\approx 0.5$ at 2 keV---c.f. Appendix~\ref{app:chandra} for sample spectra of Cas A
  at a higher spectral resolution as obtained with the Chandra ACIS-S detector.
    }
\end{figure}

In Fig.~\ref{fig:stokesI} we show the Stokes I three-band color map of Cas A with a pixel size of 10.4\arcsec, and the 3--6 keV {map of the minimal detectable polarization at 99\% confidence level} (MDP99, see Appendix~\ref{app:maps})  for a pixel size of {42\arcsec.} 
The MDP99 map shows that polarization on these pixel scales can be detected at the $\sim 5$\% polarization degree level in the interior regions, and up to 15\% in the fainter outer regions. 
All maps also show the   4--6 keV contour lines based on Chandra observations, which  serve as a point of reference for all derived polarization maps.
Fig.~\ref{fig:stokesI} includes a hardness ratio map based on the 4-6 keV and 3.4--3.6 keV maps as observed by Chandra. 
The 3.4--3.6 keV band is also relatively devoid of line emission so this hardness ratio map
shows which regions emit hard, and which regions emit soft X-ray continuum \citep[c.f. Fig.~6 in][]{helder08}. 
The  hard X-ray continuum likely corresponds to synchrotron radiation, whereas the softer continuum is dominated by thermal bremsstrahlung given that that the typical electron temperatures within
Cas A are $0.5 \lesssim T_{\rm e}\lesssim 4.5$~keV \citep[][see also Appendix~\ref{app:chandra}]{hwang12}. 
The hardness map illustrates in what regions X-ray synchrotron radiation is likely dominating the continuum emission: in the outer regions, and in some parts of the interior region---mostly the center,
but with a prominent spot in the western interior region of Cas A. See also Appendix~\ref{app:chandra}.
Although the energy resolution of IXPE is limited---with $\Delta E\approx 0.5$~keV at 2 keV and scaling inversely as the square root of the energy---one can still appreciate in the Stokes I map  (Fig.~\ref{fig:stokesI}, left)  that the western part and outer regions have harder
spectra (showing up bluish). The spectral capabilities can also be assessed from the detector 1 spectrum shown in Fig.~\ref{fig:spectra}  for three extraction regions. The Si-K and S-K line emission
do not show up as broadened lines as is the case with CCD spectra, but rather lead to noticeable inflections around 3~keV (Si-K and S-K) and around 6~keV (caused by Fe-K emission).
{For} the current paper we concentrated fully on the spatial analysis.

\begin{figure*}
  \centerline{
   \includegraphics[trim=0 0 0 4,clip=true,width=0.5\textwidth]{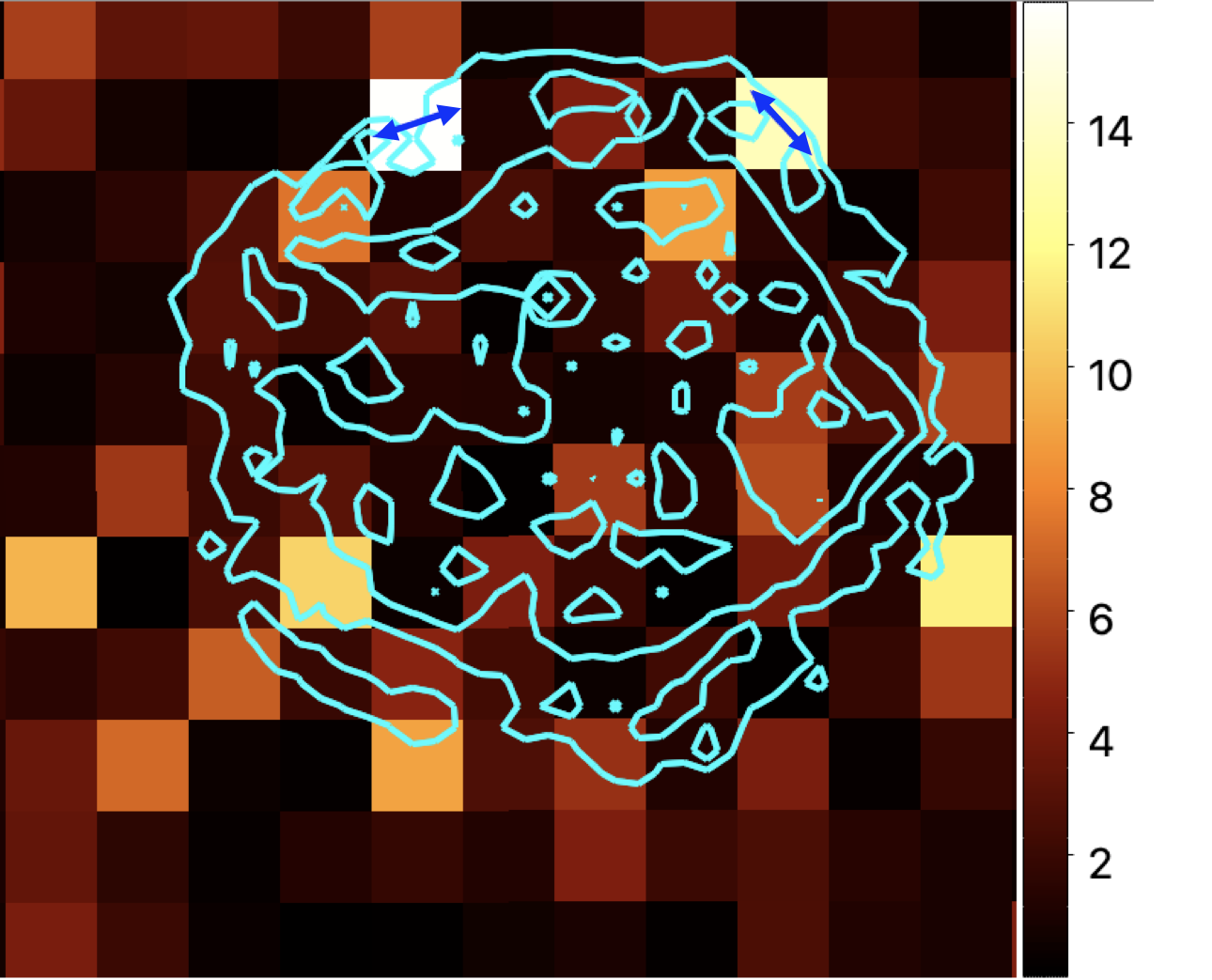}
      \includegraphics[trim=0 0 0 4,clip=true,width=0.5\textwidth]{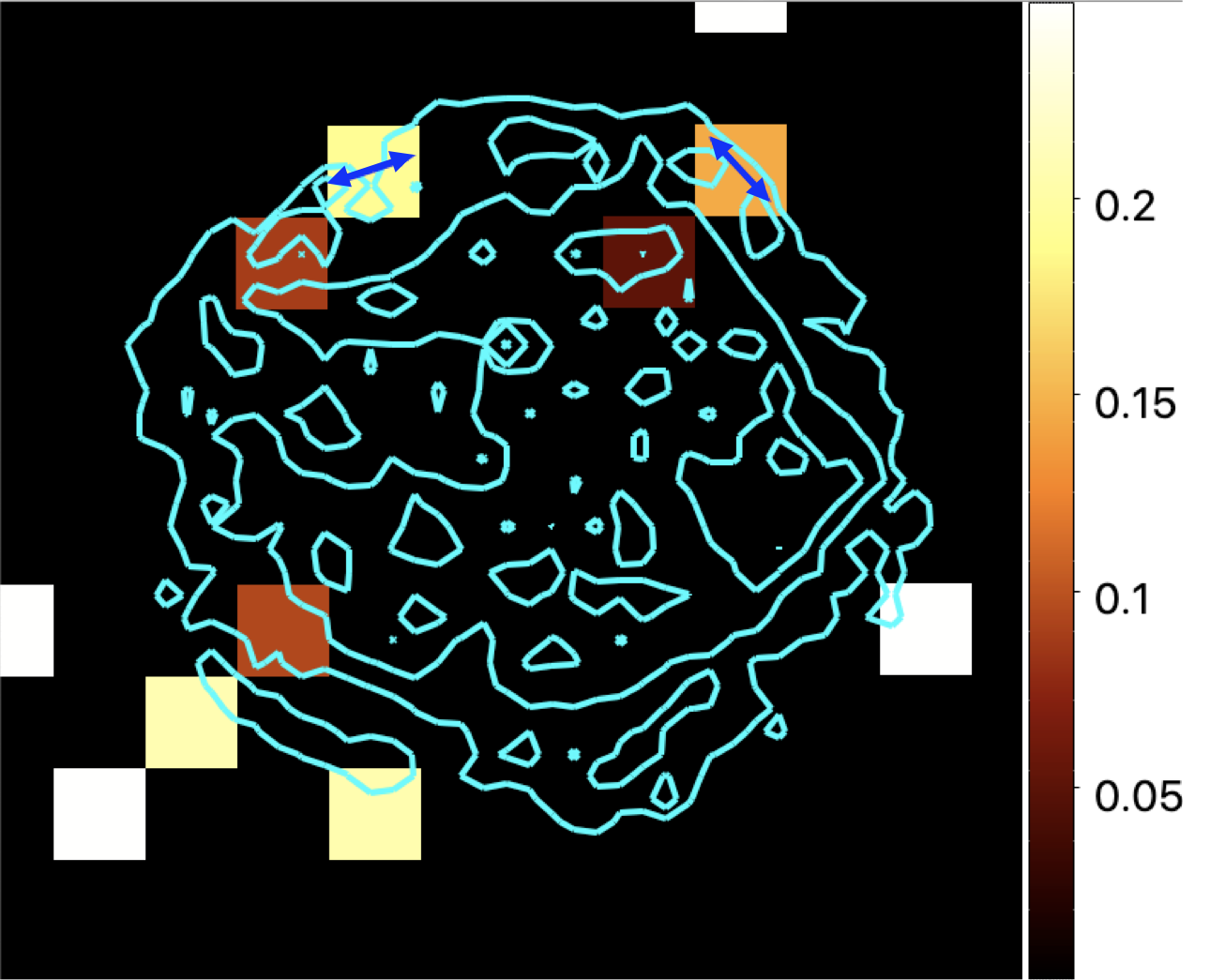}
      }
      \centerline{
      \includegraphics[trim=0 0 0 4,clip=true,width=0.5\textwidth]{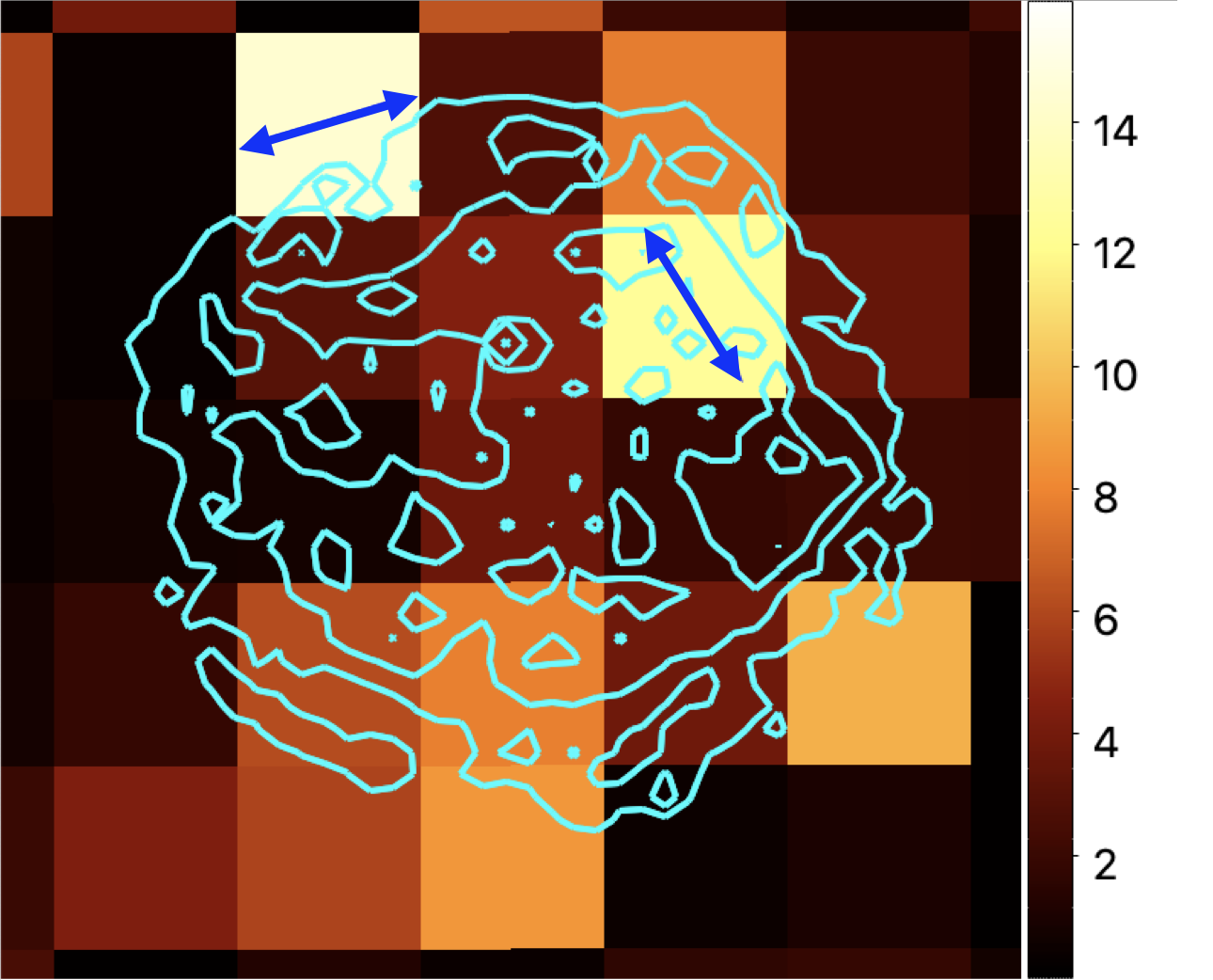}
      \includegraphics[trim=0 0 0 4,clip=true,width=0.5\textwidth]{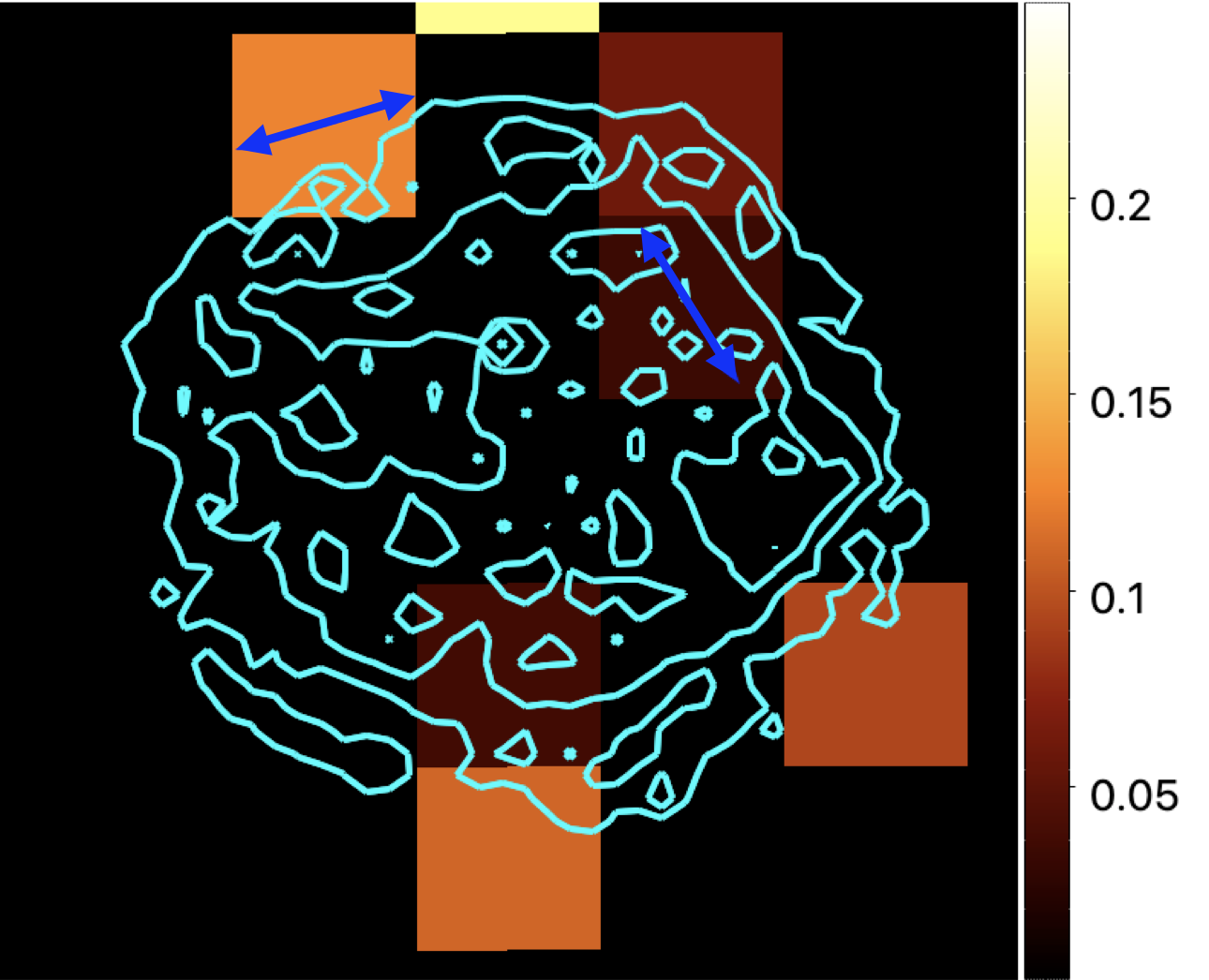}
  }
  \caption{
    \label{fig:x2pdmaps}
          {Left:} {Maps of $\chi^2_2$ (or $S_{i,j}$, see Appendix~\ref{app:maps})} values for the  polarization signal for the 3-6 keV band. 
          {Right:} the corresponding polarization degree maps. 
{Only pixels with pre-trial confidence levels above $2\sigma$ ($\chi^2_2>6.28$) are shown.}
          For pixels
          with  $\chi_2^2>11.8$ (corresponding to $3\sigma$ confidence level)  the polarization angles are indicated with blue arrows. The errors on these angles are $\sim 8^\circ$.
	Top row: maps with pixel sizes binned to 42\arcsec. Peaks in the $\chi^2_2$ map are $\chi^2_2=15.9,13.6$ corresponding to polarization degrees of
	19\%, and {14.5\%}.
	Bottom row: same plot, but with {a larger pixel size} of 84\arcsec. 
	Peaks in the $\chi^2_2$ map are $\chi^2_2=14.4,12.3$ corresponding to polarization degrees of
	12.4\% and {3.4\%}.
  }
\end{figure*}

Analyzing the full $Q$ and $U$ maps we did not find any regions for which we can report a  significant detection of a polarization signal. 
We did find regions with polarization signals above the 3$\sigma$ pre-trial confidence level.
However, given that the statistics depend on the pixel-size used, and also on the fact that Cas A is covered by about $\sim$200 angular resolution regions,
\footnote{Cas A has a diameter of $\approx$5.5\arcmin, which can be covered by $\sim$174 pixels of 25\arcsec.
}
we cannot confidently say that our analysis of the  $Q$ and $U$ maps provide  significant detections. For example, for 200 pixels the expected number of false positive detections at the 3$\sigma$ level (99.73\% confidence)
is 0.5 pixel. The post-trial probability for a {spurious} detection at the the pre-trial  3$\sigma$ level is
{ 42\% ($=1-(99.73\%)^{200}$). }
 
Nevertheless, the {  $\sim 3.5\sigma$} signals do deserve to be mentioned, first  because future, improved data analysis or additional data may be able to confirm these signals as solid detections, and secondly
because these polarization ``hotspots'' illustrate the low level of X-ray polarization degree of Cas A on small scales.

In Fig.~\ref{fig:x2pdmaps}  we show two maps of polarization significance {(using a $\chi_2^2$-map) and the polarization degree for a pixel sizes of 42\arcsec\ and 84\arcsec.} For the polarization degree map we only show the pixels with detection confidence above 99.73\% confidence level. {The most significant signal has $\chi^2_2=15.9$, corresponding to a $3.57\sigma$ confidence level.}
{The hotspots correspond to polarization degrees ranging from 3.4\% (inner region, using 84\arcsec\ pixels) to 19\% (outer, forward shock region, using 42\arcsec\ pixels)}.  These values are just above the MDP99 level. 
{In the future, with an improved analysis chain, we may be able to lower the MDP99 level and
   increase sensitivity 
   to regions with lower degrees of polarization.}

\begin{figure}
  \centerline{
    \includegraphics[trim=80 80 300 330 0,clip=true,width=0.99\columnwidth]{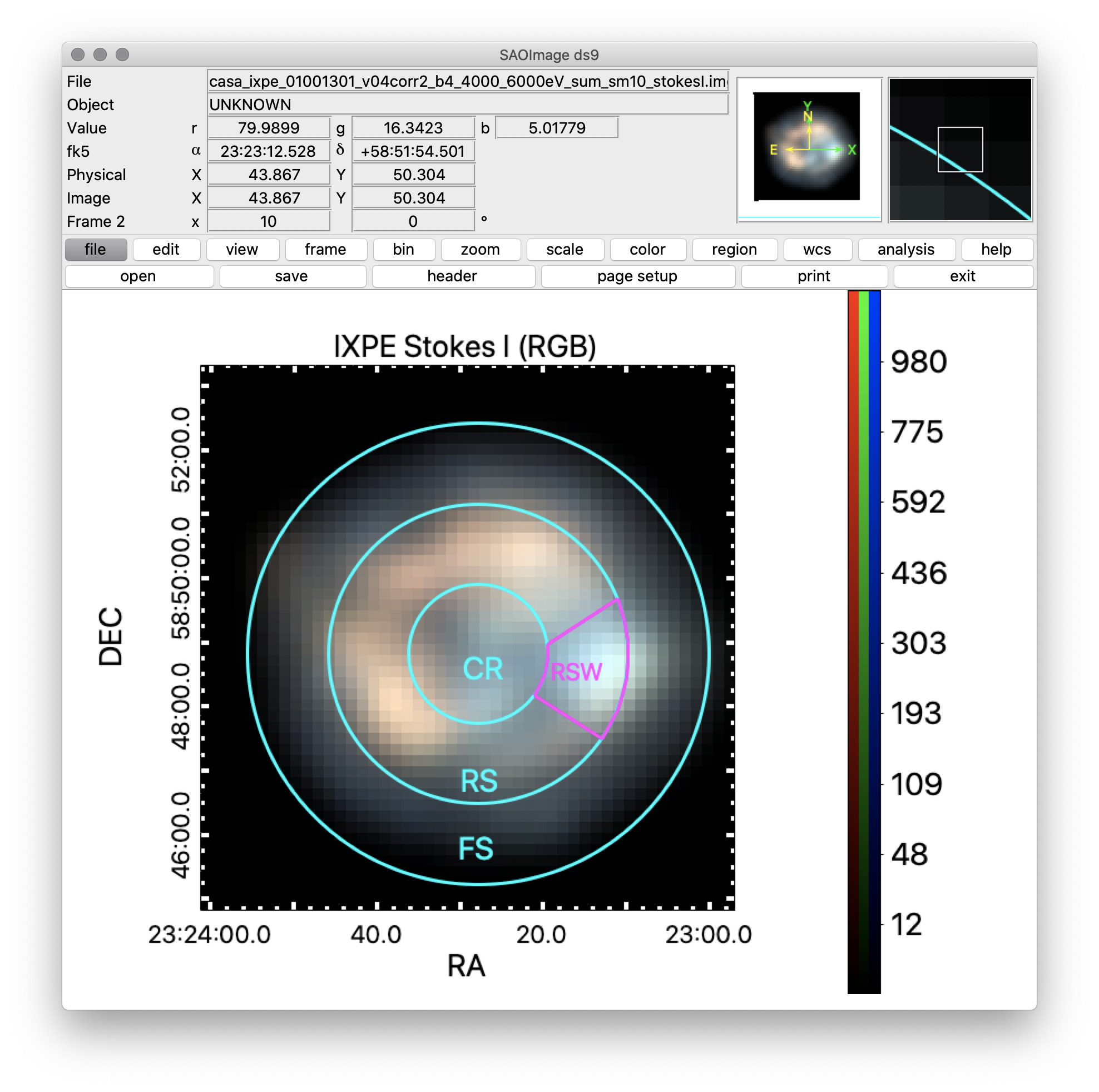}
    }
  \caption{
    \label{fig:reorientation}
    IXPE map, similar to Fig.~\ref{fig:stokesI}, with superimposed the regions that were used to test for an overall
    radial or tangential polarization vector orientation.
    See Table~\ref{tab:annuli}.
  }
\end{figure}

\begin{table*}																						
\caption{Polarization measurements in annuli and western region after imposing circular symmetry for Stokes $U$ and $Q$.\tablenotemark{a}\label{tab:annuli}	}			
\begin{tabular}{llllllcc}\hline\hline\noalign{\smallskip}																						
		&	$R_{\rm min}$\tablenotemark{b} 	&	$R_{\rm max}$	&	MDP99	&	Pol. Degree	&	PD Corrected\tablenotemark{c}	&	Angle\tablenotemark{d}		&		Significance			\\		
		&	(arscec)	&	(arcsec)	&	(\%)	&	(\%)	&	(\%)	&	($^\circ$)		&					\\\noalign{\smallskip}\hline\noalign{\smallskip}		
{Central region} (CR)		&	0	&	65	&	2.4	&	$<3.1$	&	$<3.7$	&	N/A		&	$	0.9	\sigma	$	\\		
Reverse shock (RS)		&	65	&	140	&	1.3	&	$1.6 \pm 0.4$	&	$2.2\pm 0.6$	&	$77.2\pm 7.6$		&	$	3.1	\sigma	$	\\		
RS West (RSW)	\tablenotemark{e}	&	65	&	140	&	2.6	&	$<3.9$	&	$<5.1$	&	N/A		&	$	1.9	\sigma	$	\\		
Forward shock (FS)		&	140	&	216	&	2.3	&	$3.5 \pm 0.7$	&	$4.5 \pm 1.0$	&	$89.8\pm 6.1$		&	$	4.1	\sigma	$	\\		
FS + RSW	\tablenotemark{f}	&	216	&	216	&	1.7	&	$3.0 \pm 0.6$	&	$3.8\pm 0.7$	&	$87.2\pm 5.4$		&	$	4.8	\sigma	$	\\		
All		&	0	&	216	&	1.0	&	$1.8 \pm 0.3$	&	$2.5\pm 0.5$	&	$85.7\pm 5.2$		&	$	4.9	\sigma	$	\\\noalign{\smallskip}\hline\noalign{\smallskip}		
\end{tabular}		
\tablenotetext{a}{See main text for an explanation. {Listed errors correspond to $1\sigma$ (68\% confidence) ranges.}}	
\tablenotetext{b}{The central coordinate used was RA$_{J2000}=23^{\rm h}23^{\rm m}27.8^{\rm s}$, DEC$_{J2000}=58^\circ48\arcmin49.4\arcsec$, which is the explosion center determined by \citet{thorstensen01}.}	
\tablenotetext{c}{The inferred polarization degree for the synchrotron component only, see Appendix~\ref{app:chandra}}
\tablenotetext{d}{An angle of 90$^\circ$ corresponds to tangential polarization vector associated with a radial magnetic field orientation.}
\tablenotetext{e}{This region is a subset of the "Reverse shock" region.}															
\tablenotetext{f}{This combines the Reverse shock West and Forward shock region.}															
\end{table*}				

\subsection{Analysis based on an assumed circular magnetic-field topology}
\label{sec:aligned}

The pixel-by-pixel polarization measurement yielded tentative detections of polarization with polarization degrees of 4\%--15\% at the 3$\sigma$--4$\sigma$ confidence level, too low
to claim a solid detection.
Binning into larger pixel sizes {improves} the polarization statistics, but at the expense of potential depolarization  due to  the mixing of regions with different polarization angles. 
However, given the roughly spherical symmetry of Cas A itself, as well as the long-known radial symmetry of the magnetic-field orientation as inferred from
radio observations, one can improve the statistics by summing over large regions by assuming a circular symmetry to the polarization direction.
Although the radio polarization measurements {suggest} an a priori radial magnetic-field orientation, shock compression of a highly turbulent magnetic field
instead would lead to an enhancement of the tangential component \citep[e.g.][]{jun96,bykov20}. 
Note that the polarization direction for synchrotron radiation is perpendicular to the magnetic-field orientation: a tangential polarization signal corresponds to a radial magnetic field, 
and vice versa.

Assuming a circular symmetry for the polarization direction, we 
recalculated the $q_k$ and $u_k$ values for each event, 
{by calculating  a new zero for the direction of the photo-electron ($\phi$)
based on the  
sky coordinate, and its position angle  with respect to the center of Cas A, for which we used the explosion center determined by \citet{thorstensen01}.}
This procedure results in new values { $q_k^\prime$ and $u_k^\prime$, which}
 can be summed over large regions to provide an overall signal corresponding to the radial and tangential Stokes parameters
 $Q^\prime$ and Stokes $U^\prime$.

In Fig.~\ref{fig:reorientation} we show the chosen annuli for which we obtained  $Q^\prime$ and $U^\prime$. 
They cover the central region, the reverse shock regions (overlapping with the bright shell), and the outer region, which contains the synchrotron filaments associated with the forward shock. 
Since the western reverse shock region shows evidence for strong X-ray synchrotron emission {\citep[e.g., see][and also the bluish part in the multiband Stokes I image in Fig.~\ref{fig:stokesI}]{helder08,grefenstette15},}
 we also isolate just the western  part of the reverse shock region.
 
 The resulting polarization measurements are listed in Table~\ref{tab:annuli}.  Apart from the polarization degree we list under ``PD Corrected’’, the inferred polarization degree of the synchrotron component only. 
The correction factors are based on a full modeling of the expected IXPE data by folding the Chandra best-fit spectral models through the IXPE spectral and spatial response functions using \texttt{ixpeobsim}.
The Chandra best-fit spectral models and their implications for the nonthermal flux fraction are detailed in Appendix~\ref{app:chandra}.
We stress here that the corrected PD is model-dependent as it is based on spectral fitting results. Nevertheless, the finding reported in the Appendix that up to 99\% of the continuum flux is synchrotron radiation is supported
by NuSTAR observations showing that an extrapolation of the hard X-ray (synchrotron) spectrum extrapolated to the 4--6 keV band accounts for a dominant fraction of the continuum flux
\citep[see Fig. 5 in][]{grefenstette15}.

In Appendix~\ref{app:bandeffects} we show the breakdown of the polarization signals for the ``All” and FS+RSW” regions in the 3--4 keV and 4--6 keV band, which shows that
even for the nearly lineless 4--6 keV band the measured polarization degree is indeed low, although measured at a lower significance.

\begin{figure*}
  \centerline{
    \includegraphics[trim=10 0 62 0,clip=true,height=0.267\textwidth]{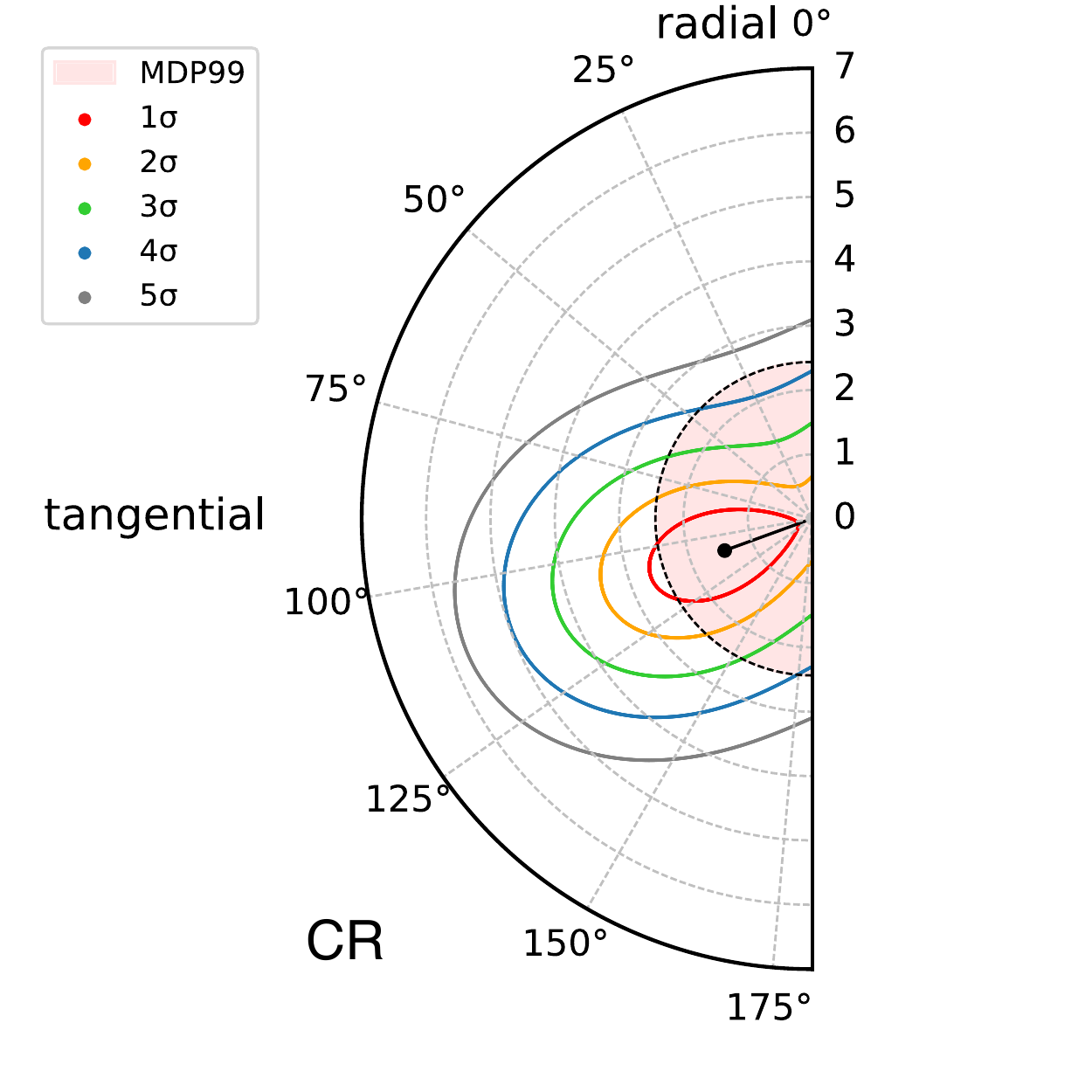}
    \includegraphics[trim=95 0 65 0,clip=true,height=0.267\textwidth]{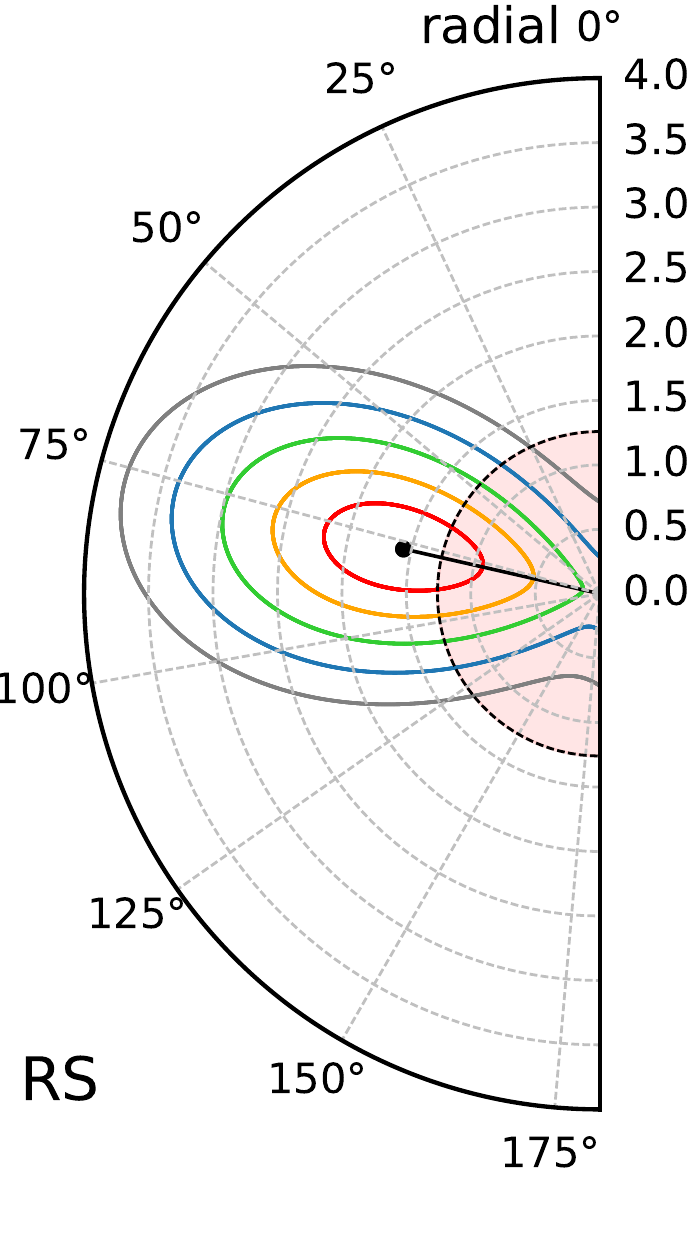}
    \includegraphics[trim=95 0 65 0,clip=true,height=0.267\textwidth]{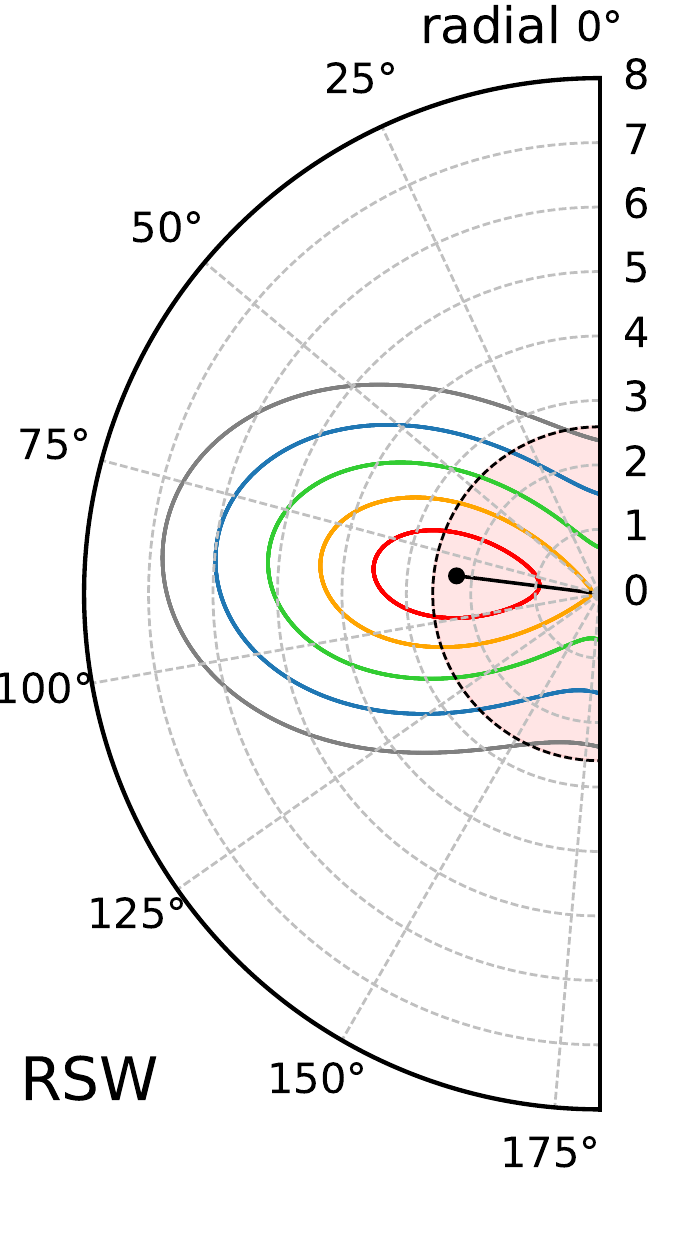}
    \includegraphics[trim=95 0 65 0,clip=true,height=0.267\textwidth]{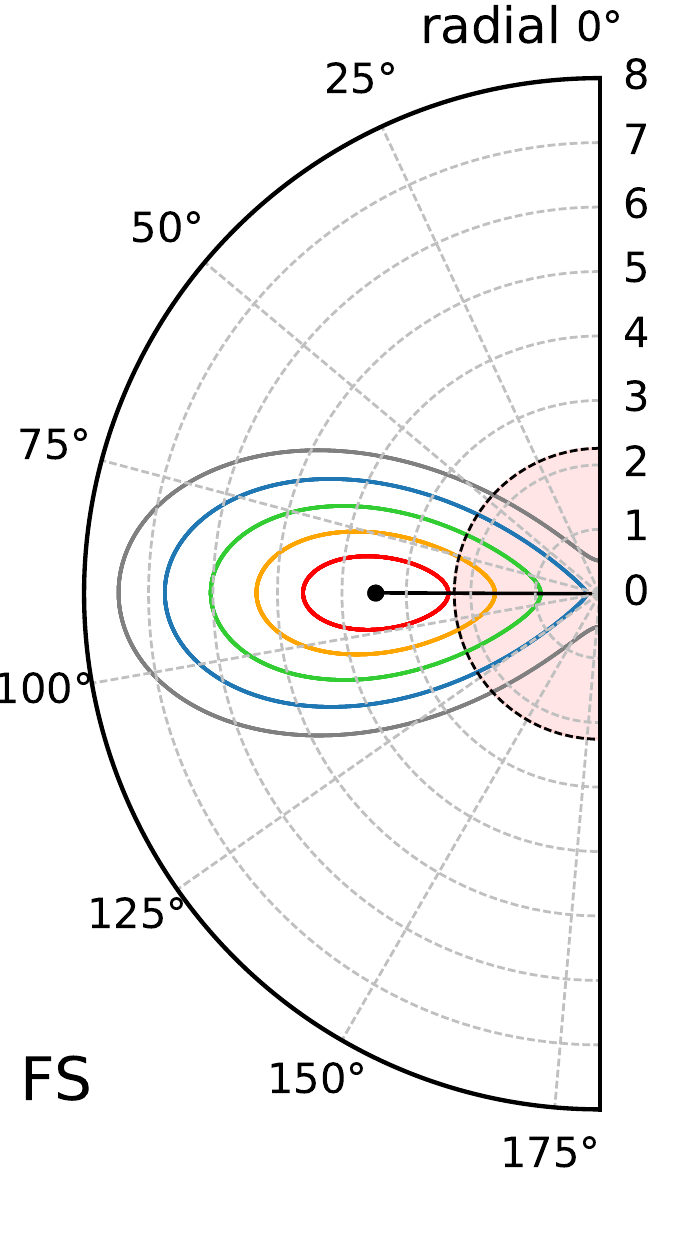}
    \includegraphics[trim=95 0 65 0,clip=true,height=0.267\textwidth]{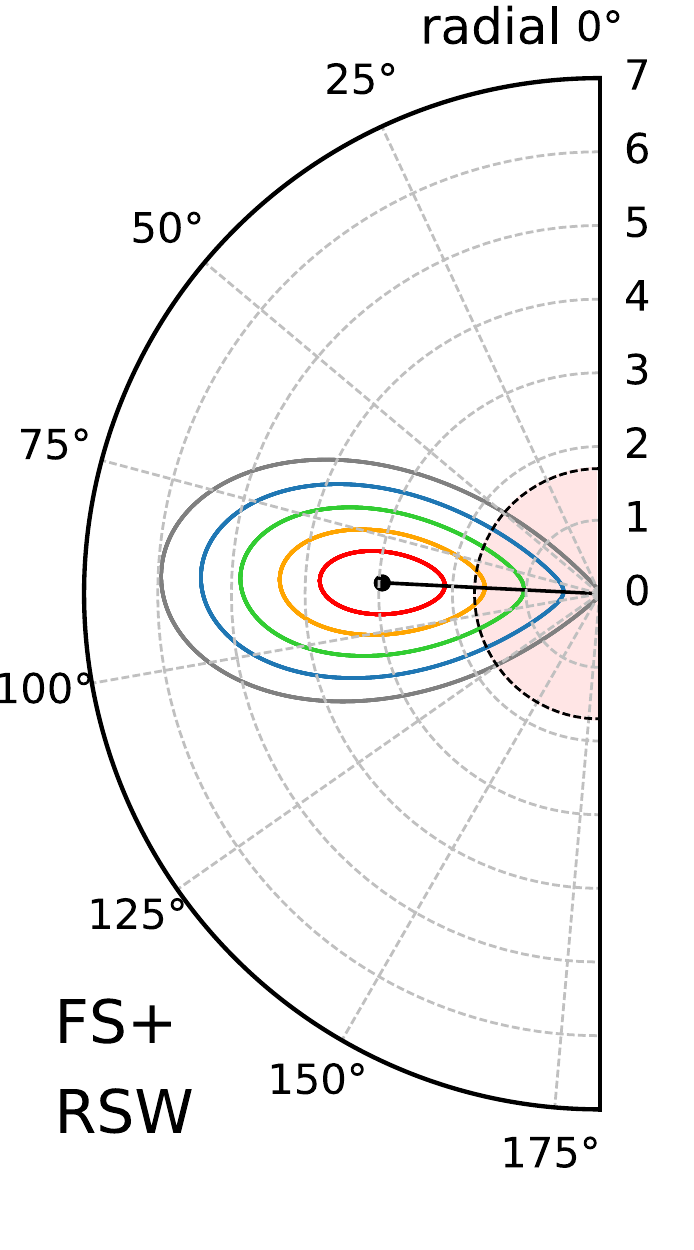}
    \includegraphics[trim=95 0 40 0,clip=true,height=0.267\textwidth]{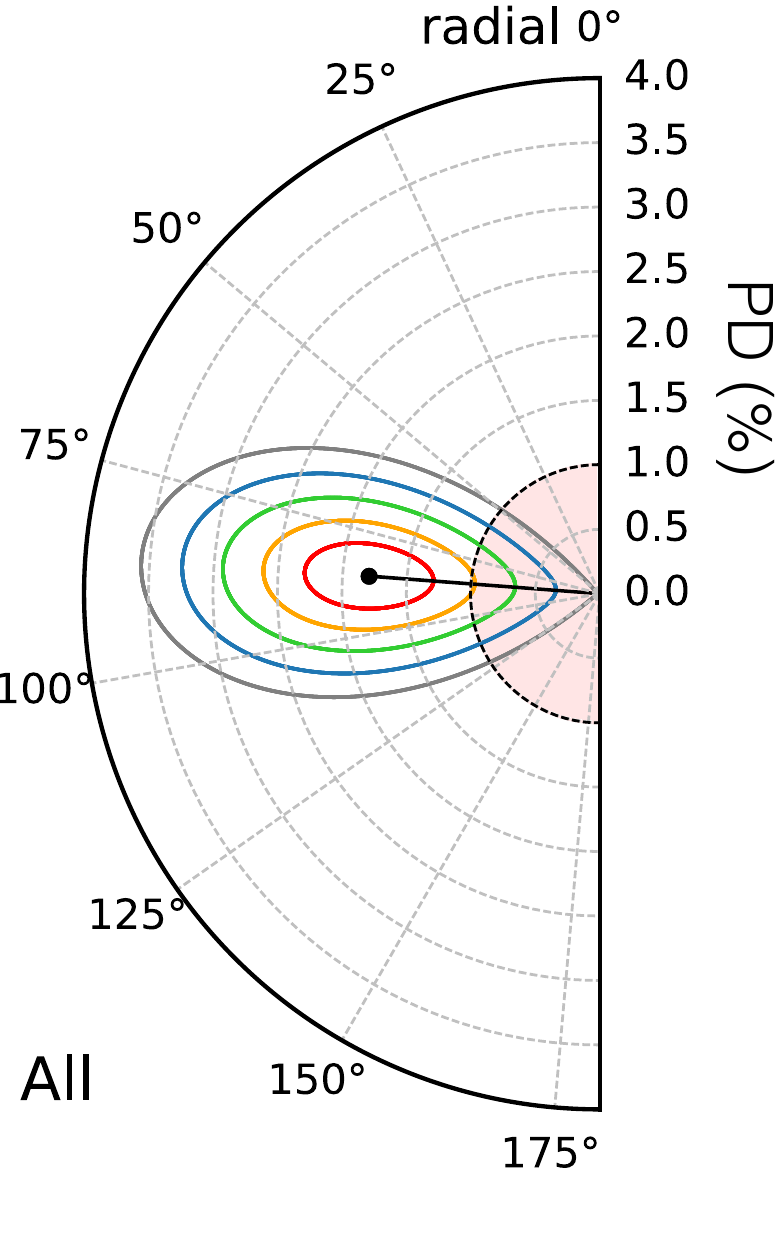}
    }
  \caption{\label{fig:polar}
    Polar diagrams depicting the {measured polarization degree and angle with respect to circular symmetry} as confidence contours  
    for  the six regions listed in Table~\ref{tab:annuli}.
      The radial coordinate indicates the polarization degree in percent. The pink region corresponds to the MDP99 level.
       Values around 90\degr\ correspond to an overal tangentially oriented polarization  averaged over the region, while around 0\degr\  indicates on average a radially oriented polarization.
  }
\end{figure*}

The errors listed should be used with caution, since polarization degree and polarization angle are not
independent statistical quantities, both being derived from a combination of $Q^\prime$ and $U^\prime$. A better representation of the measurements is obtained 
by graphically representing the Stokes $Q^\prime$ and $U^\prime$ parameters in a polar diagram, with
confidence contours.
These are plotted in Fig.~\ref{fig:polar}, which shows that for the ``forward shock region'' and the ``forward shock + western region'' the $1\sigma$ confidence contour is outside the MDP99 region (in pink).
For the  ``forward shock region" the polarization detection significance is over $4\sigma$ pre-trial, corresponding to a roughly $3\sigma$ post-trial confidence level, while for
the  ``forward shock + west region" and all data combined the detection is close to  a {$5\sigma$ pre-trial confidence level,
  corresponding to a better than $4\sigma$ post-trial confidence level}. Pre-trial refers here to the confidence level for polarization detection for a given region, whereas post-trial corrects for
  the fact that we are attempting to measure polarization for three independent regions.
Note that the underlying statistics
are not Gaussian, so the quoted  $\sigma$ values are for ease of interpretation, whereas the diagrams show the full confidence areas {in the polarization-degree versus polarization-angle plane.}
For the most significant detections we find that the polarization angle is consistent with a tangential polarization pattern, corresponding to a radial magnetic-field orientation.

\section{Discussion}
\label{sec:discussion}

The IXPE observations of Cas A reveal that the polarization level is  low {at the spatial resolution of IXPE ($\sim 30$\arcsec)}, with a tangentially oriented polarization component with a polarization degree of {
 $\approx 1.5$ $\sim 3.5$\%, corresponding to  $\approx 2.5$  to $\sim 5$\%} for the synchrotron component only.  
The polarization signal  is associated with the X-ray synchrotron bright features in Cas A: the outside region associated with the forward shock, and the western part of the reverse
shock region \citep[see][for maps of the X-ray synchrotron radiation]{vink03a,helder08,uchiyama08,grefenstette15}.
The detection of the weak polarization signal requires the assumption of circular symmetry, 
but even without this assumption the polarization degree cannot be much larger than 4\% for the interior (reverse shock) of Cas A,
and 15\% at the outer boundary, as indicated by the MDP99 levels, and the tentative indications of polarization for individual regions with angular extents of $\sim$42\arcsec--84\arcsec. 
High degrees of polarization may be present for unresolved regions
and they need to have more or less random polarization directions.

The low polarization degree and the tangential polarization pattern are compatible to what has been measured in the radio band
 \citep[e.g.][]{rosenberg70b,braun87b,anderson95b}. 
 Nevertheless, these X-ray polarization results still come as a surprise.
First of all, in X-rays the synchrotron spectrum has steepened, leading
to higher intrinsic polarization degrees \citep{ginzburg67,bykov09}. Secondly,  X-ray synchrotron radiation in Cas A originates from a narrow region of $\lesssim 10^{17}$~cm immediately
downstream of the shocks, which reduces depolarization caused by the line of sight crossing regions with different polarization angles. And thirdly, at the shock front the perpendicular (tangential) component
of the magnetic field is compressed, so close to the shock front a radial instead of a tangential polarization angle is expected.

Apart from line of sight effects, a low polarization degree may indicate a turbulent magnetic field, in particular for angular scales smaller than the IXPE angular resolution.
The level of magnetic-field turbulence near the shock regions in Cas A is expected to be high, as a high level of turbulence is required for accelerating electrons to $\gtrsim 10$~TeV --- the energy
needed for producing X-ray synchrotron radiation. However, efficient DSA assumes a high magnetic-field turbulence in the unshocked plasma, whereas the shock compression
should enhance the tangential component of the magnetic field.

The origin of the magnetic-field turbulence necessary for DSA is often attributed to the interactions of cosmic rays in the precursor with the preshock plasma.
These interactions are divided into resonant and non-resonant cosmic-ray interactions  \citep[see][for a review]{marcowith16}. For the resonant interactions
the energetic charged particles interact and amplify Alfvén waves with  wavelengths comparable to the  gyroradii of the cosmic-ray particles.
A prime example of a non-resonant  interaction is the Bell streaming instability \citep{bell04}.
It is caused by the response of the plasma to the electric current driven into the plasma by the cosmic rays.
Apart from cosmic-ray induced magnetic-field turbulence, which is expected to affect the pre-shock magnetic turbulence, the shock itself may also induce post-shock turbulence.
For example, recent MHD simulations by \citet{hu22} show substantial post-shock magnetic-field turbulence, which may be contributing to the low polarization degree in the X-ray synchrotron filaments.
Moreover, collisionless shocks in the heliosphere are known to be rippled, rather than smooth \citep[see for example][and references therein]{gedalin22}. If these ripples are also present for the higher Mach number
collissionless shock in Cas A, they  may also affect its postshock magnetic-field turbulence.

Simulations of  X-ray polarization signals to be expected for young SNRs by IXPE by \citet{bykov20} predicted polarization degrees of 30\%--70\%, assuming an anisotropic magnetic field---i.e.
with a preferred orientation due to shock compression---and with a turbulence spectrum cutting off at $10^{18}$~cm, which is compatible with the IXPE angular resolution.
For isotropic magnetic fields the polarization degree  predicted  by \citet{bykov20} is similar
to the values reported here: $\sim 5\% -15\%$.  Based on these models the magnetic-field turbulence near the shock may be close to isotropic, perhaps due to nonlinear 
interactions of the fluctuations downstream from the shock, as hypothesised by \citet{bykov20}. 
In addition, it may also indicate that the  magnetic-field turbulence wavelength cutoff is well
below the cutoff of $10^{18}$~cm assumed for the simulations by \citet{bykov20}.

Indeed, the theoretical expectations for the maximum turbulence wavelength suggest values well below $10^{18}$~cm.
For the resonant instability one needs to consider the gyroradius of the particles, which is $r_{\rm g}=3\times 10^{15} E_{14}B_{-4}^{-1}~{\rm cm}$. Even if protons are accelerated up to $10^{15}$~eV---for which there
is no evidence for the case of Cas A \citep{magic17_casa} --- we expect that the turbulence cuts off at length scales of $<10^{16}$~cm, much smaller than the IXPE pixel size, and also smaller than the thickness of the X-ray synchrotron emitting
region of $\sim 10^{17}$~cm.

For the longest lengthscales caused by the nonresonant Bell instability we refer to Eq.~21 in \citet{bell04}. This equation predicts for  $V_{\rm sh}=5500~{\rm km\,s^{-1}}$ a
maximum length scale for  magnetic-field turbulence of $l_{\rm Bell}\approx 8\times 10^{16}E_{{\rm max},14}(B_0/3~{\rm \mu G})$~cm.  
$B_0$ refers to the pre-amplified parallel magnetic-field component, and $E_{{\rm max},14}$ to the maximum energy of the cosmic rays in units of $10^{14}$~eV.
This shows that even for $E_{\rm max}=10^{15}$~eV the maximum turbulence scale for the Bell instability is less than $10^{17}$~cm--- too small for IXPE to detect
 peaks in the polarization signal caused by the largest scale turbulence modes.

An alternative reason for the low degree of polarization may be that the synchrotron emission is coming from a mix of  plasma with the expected tangential oriented magnetic field immediately downstream of
the shock, and a more radially oriented magnetic field further downstream of the shock, but still within $10^{17}$~cm of the shock.

Although a radial magnetic-field orientation throughout the SNR  has been inferred from radio observations, its origin has never been fully understood.
There are two main hypotheses: (i) there are  magnetohydrodynamical (MHD) process stretching the magnetic fields, and (ii) the radial magnetic field is
the result of 
a {selection effect},
related to efficient acceleration wherever the magnetic-field orientation at the shock {happens to be} quasi-parallel.

As for hypothesis (i),
it was long suspected that the radially oriented magnetic fields are caused by filamentation due hydrodynamical instabilities,  such as Rayleigh-Taylor instabilities at the contact
discontinuity between shocked ejecta and shocked circumstellar plasma \citep{gull73}.  MHD simulations confirm the formation of a radially oriented magnetic field
near the contact discontinuity, but the regions near the forward shock are predicted to have a tangential magnetic field \citet{jun96}.
\citet{inoue13} performed MHD  simulations of the shock front in the presence of a clumpy medium, resulting in Richtmeyer-Meshkov instabilities
which reorient the overall magnetic-field orientation from a tangential to a radial direction within a distance of
$\sim 7\times 10^{17}$~cm from the shock.
This length scale is almost an order of magnitude
larger than the width of the X-ray synchrotron filaments, but the MHD length scale is dependent on environmental conditions, such as the size and density contrasts of the clumps.
{Very Large Array} (VLA) radio polarization measurements by \citet{gotthelf01a} lend some support to this hypothesis, as they report a swing in polarization angle near the forward shock.
However, the radio polarization measurements were based on an azimuthal integration over $55^\circ$, {resulting in an effective
spatial resolution that is larger than the widths of the X-ray filaments.}

 Regarding hypothesis (ii), 
\citet{west17} investigated what the impact is if DSA preferentially occurs wherever the shock has a quasi-parallel magnetic-field orientation
 (i.e., a significant component parallel to the shock velocity vector),
motivated by the likely magnetic-field orientation in SN\,1006 \citep{rothenflug04}. As the plasma is advected downstream and keeps roughly its magnetic-field orientation,
the synchrotron radiation becomes biased toward radial magnetic fields.
The likely reason for efficient electron acceleration at quasi-parallel shocks probably concerns the initial injection of low energy electrons into the acceleration process.
If the observational selection bias still pertains to X-ray synchrotron radiation, the implication is that 
 the  responsible  $\gtrsim 10$~TeV electrons should not have diffused too far away from the original regions with preferentially radial magnetic fields. 
 This would imply that the shock regions with quasi-parallel magnetic fields are associated with a characteristic turbulence wavelength, $\lambda_B$, that must be comparable in size to
 the diffusion length scale ($l_{\rm diff}$).  For $l_{\rm diff}\ll \lambda_B$ one expects a strong polarization signature, whereas for  
 $l_{\rm diff}\gg \lambda_B$ the 
 {selection effect}
 hypothesis by \citet{west17} should not pertain to X-ray synchrotron radiation.
 The relevant diffusion length scale is that for the tangential diffusion component, which should be
 $l_{\rm diff}\approx \sqrt{D\tau_{\rm loss}}\approx 3\times 10^{16}\eta^{1/2}(B/250~{\rm \mu G})^{-1/2}$~cm, with the diffusion
 coefficient given by $D=\frac{1}{3}\eta cE/(eB)$, {where $\eta\geq 1$ is the ratio between the mean-free path of the particles  and their gyroradius  (see Introduction for the other definitions). }
 The length scale of $3\times 10^{16}$~cm corresponds to 0.6\arcsec, which  is comparable to the Chandra point spread function. 
 Since the X-ray synchrotron filaments as seen by Chandra appear to be smooth, hypothesis (ii) requires the
 diffusion length scale to be  smaller than $3\times 10^{16}$~cm, requiring $\eta \sim 1$ and $B>250~{\rm \mu G}$.
 
 So although the IXPE observations do not favor one hypothesis for the radial magnetic-field orientation over the other, the measured polarization signal 
puts constraints on relevant
length scales. For the hypothesis that an MHD process leads to  intrinsically radial magnetic-field alignments---rather than through a perseived radial magnetic field \citep{west17}--- the IXPE results
require that the reorientation happens within  $10^{17}$~cm downstream of the shock. In contrast, the hypothesis that the radial magnetic field orientation is caused by
{a selection effect}
sets constraints on the wavelength of the magnetic-field turbulence of $\lambda_B \lesssim 3\times 10^{16}$~cm.
 
\section{Conclusions}
\label{sec:conclusions}

We reported here on the very first detection of X-ray polarization from a shell-type SNR---the young and bright core-collapse SNR Cas A, which was the first science target of the NASA IXPE mission.
We employed two different methods to measure the  polarization  signals: a pixel-by-pixel analysis, and a more sensitive analysis that assumes a circular symmetry for the polarization vectors,
and then summing over large regions. 

The pixel-by-pixel analysis provides tentative hints of polarization in the 3--6 keV band at the 3$\sigma$ to 4$\sigma$ pre-trial confidence level, with associated polarization degrees ranging from 3\% to 19\%, just above the local minimum detectable
polarization degree at 99\% confidence (MDP99). {Taking into account post-trial factors these detections are not significant.}
The  analysis assuming circular symmetry for the polarization angles  provides solid detections (close to $5\sigma$) for an annular regions encompassing  the forward shock region, and a region
covering the  entire SNR. 
The corresponding polarization degrees are lower than the polarization generally reported in the radio: 1.8--3.5\% in X-rays versus $\sim 5$\% in the radio band { \citep[e.g., see][for radio polarization measurements with a resolution comparable to IXPE]{rosenberg70b}}. Even after correction for the contribution of thermal X-ray 
emission in the 3-6 keV band, this suggests  an X-ray synchrotron polarization degree of $\sim 2$--5\%, which is similar, or slightly lower than in the radio band. Like in the radio band, the polarization vectors suggest an overall radial magnetic-field orientation.

Since the X-ray synchrotron radiation is confined to regions within $\lesssim 10^{17}$~cm, the IXPE results imply that the radial magnetic-field structure is already present close to the shock, which puts constraints on the physical processes that leads to the radial magnetic-field structures measured in young SNRs.

The measured low polarization degree
is consistent with a nearly isotropic magnetic-field turbulence as simulated
by \citet{bykov20},  but it could also be caused by a mixture of tangential magnetic-field alignment close to the shock front, and a radially alligned magnetic field structure further downstream of the shock.

\begin{acknowledgments}
The Imaging X ray Polarimetry Explorer (IXPE) is a joint US and Italian mission.  The US contribution is supported by the National Aeronautics and Space Administration (NASA) and led and managed by its Marshall Space Flight Center (MSFC), with industry partner Ball Aerospace (contract NNM15AA18C).  The Italian contribution is supported by the Italian Space Agency (Agenzia Spaziale Italiana, ASI) through contract ASI-OHBI-2017-12-I.0, agreements ASI-INAF-2017-12-H0 and ASI-INFN-2017.13-H0, and its Space Science Data Center (SSDC), and by the Istituto Nazionale di Astrofisica (INAF) and the Istituto Nazionale di Fisica Nucleare (INFN) in Italy.  This research used data products provided by the IXPE Team (MSFC, SSDC, INAF, and INFN) and distributed with additional software tools by the High-Energy Astrophysics Science Archive Research Center (HEASARC), at NASA Goddard Space Flight Center (GSFC).
J.V. \& D.P. are supported by funding from the European Union's Horizon 2020 research and innovation program under grant agreement No. 101004131 (SHARP).
{
The research at Boston University was supported in part by National Science Foundation grant AST-2108622.
We thank Dawoon Kim for kindly providing us with his script for making polar plots of polarization degree and angle.}
\end{acknowledgments}

\vspace{5mm}
\facilities{IXPE}

\software{
\texttt{ixpeobsim} \citep{baldini22}
          }


\iftrue

\fi

\appendix

\section{Calculation of the $Q$ and $U$ maps}
\label{app:maps}

The polarization of radiation as measured by IXPE is calculated based on the reconstructed ejection angle of the photo-electron, $\phi_k$, associated with event $k$.
Maps of the Stokes parameters $I$, $Q$ and $U$ can then be obtained using (weighted) sums of $q_k=2\cos 2\phi_k$ and $u_k=2\sin 2\phi_k$, which can also include  the correction for the energy- or event-dependent modulation factor $\mu_k$:
\begin{align}
I_{i,j} =  & W_{i,j}^{-1}\sum_k \frac{w_{k,i,j}}{\mu_k}, \\ \nonumber
Q_{i,j} =  &  W_{i,j}^{-1} \sum_k \frac{w_{k,i,j} q_{k,i,j} }{\mu_k},    \\ \nonumber
U_{i,j} =   & W_{i,j}^{-1} \sum_k \frac{w_{k,i,j} u_{k,i,j} }{\mu_k},
\end{align}
with $W_{i,j}\equiv \sum_k w_{k,i,j}$ the sum of the weights. This definition of Stokes $I$ is chosen so that the polarization fraction can be expressed as $\Pi_{i,j} = \sqrt{Q_{i,j}^2+U_{i,j}^2}/I_{i,j}$. Note that the division by $\mu_k$ ensures
that the detector response to polarization, which is energy dependent, is taken into account.

The weights $w_k$ can be based on the quality of the reconstruction of $\phi_k$  \citep{marshall21,dimarco22}, or  on the energy dependent modulation factor \citep{vink18a}, $w_k=\mu_k^{-2}$, or simply ignored, i.e. $w_k=1$. The variances in the Stokes parameters are
\begin{align}
\Var(I_{i,j})=  & W_{i,j}^{-2}\sum_k \frac{w_{k,i,j}^2}{\mu_k^2}, \\ \nonumber
\Var(Q_{i,j}) =  &  W_{i,j}^{-2} \sum_k \frac{w_{k,i,j}^2 q_{k,i,j}^2 }{\mu_k^2},    \\ \nonumber
\Var(U_{i,j}) =   & W_{i,j}^{-2} \sum_k \frac{w_{k,i,j}^2 u_{k,i,j}^2}{\mu_k^2}.
\end{align}

The polarization detection significance can be obtained by the methods described in e.g. \citet{kislat15,vink18a}, but we note here that in the absence of polarization the expectation values for the Stokes parameters
are $E[Q]=0$ and $E[U]=0$, and that  $\cos 2\phi$ and $\sin 2\phi$ are orthogonal functions. As a result the statistical quantity
\begin{equation}
S_{i,j} \equiv \frac{Q_{i,j}^2}{\Var(Q_{i,j})} + \frac{U_{i,j}^2}{\Var(U_{i,j})}
\end{equation}
has a $\chi^2$ distribution with two degrees of freedom. 
{One can make a map of $S_{i,j}$, which we call a $\chi^2$ map, that 
shows how consistent the observed polarization signal is with the null hypothesis (i.e., no polarization signal) in pixel ($i,j$).
A large value of $S_{i,j}/\chi_{2,i,j}^2$ indicates, therefore, the presence of a  polarization signal for pixel $i,j$ with a confidence corresponding to $\chi^2_2$. 
A 3$\sigma$ signal corresponds to $\chi^2_2=11.8$, $4\sigma$ to  $\chi^2_2=19.3$, and $5\sigma$ to  $\chi^2_2=28.7$. }

In the absence of an intrinsic polarization signal the expectation values for the variances are $E[\Var(Q_{i,j})]=E[\Var(U_{i,j})]=2W_{i,j}^{-2}\sum_k (w_{k,i,j}^2/\mu_k^2)$,
since $E[(u^2+q^2)]=4$.
Assuming that  $\Var(Q_{i,j}) \approx \Var(U_{i,j})$, we can express the polarization degree
as
\begin{equation}
\Pi_{i,j} = \sqrt{ \frac{Q_{i,j}^2+U_{i,j}^2}{I_{i,j}^2}} = \sqrt{ \frac{S_{i,j}\Var(Q_{i,j})}{I_{i,j}^2}}.
\end{equation}
By noting that $S_{i,j}$ is $\chi_2^2$  distributed and that the 99\% confidence range corresponds to $\chi_2^2=9.21$ we see that the minimal detectable polarization (MPD$_{99}$) degree
is 
\begin{equation}
{\rm MDP}_{99}=  \sqrt{ \frac{9.21 \times 2W_{i,j}^2\sum_k (w_{k,i,j}^2/\mu_k^2)}{W_{i,j}^{-2}\sum_k (w_{k,i,j}^2/\mu_k)^2}}= \frac{4.29}{\sqrt{\sum_k (w_{k,i,j}^2/\mu_k)}},
\end{equation}
equivalent to the expression derived by \citet{kislat15}. 
The extraction of Stokes parameters over larger regions, after a local rotation of $q_k$ and $u_k$, follows the same procedure as outlined here, except that the final
outcome should not have a pixel index $(i,j)$, but a region label.

Note that summing together $S_{i,j}$ over various pixels, by for example rebinning ($S_{i^\prime,j^\prime} =\sum_{i,j}S_{i,j}$), results in yet another $\chi^2$ distribution, but now with
more degrees of freedom. For example, rebinning $S_{i,j}$ by $2\times2$ pixels, results in $\chi^2_8$ distribution. Essentially this is an incoherent addition of the signal,
so less sensitive than rebinning $Q_{i,j}, U_{i,j}$ (i.e. coherent summing), but it has the advantage that it is not sensitive to rotations of the polarization vector within the binned pixels region.

\begin{figure}
          \centerline{
\includegraphics[trim=55 100 60 263,clip=true,width=0.33\textwidth]{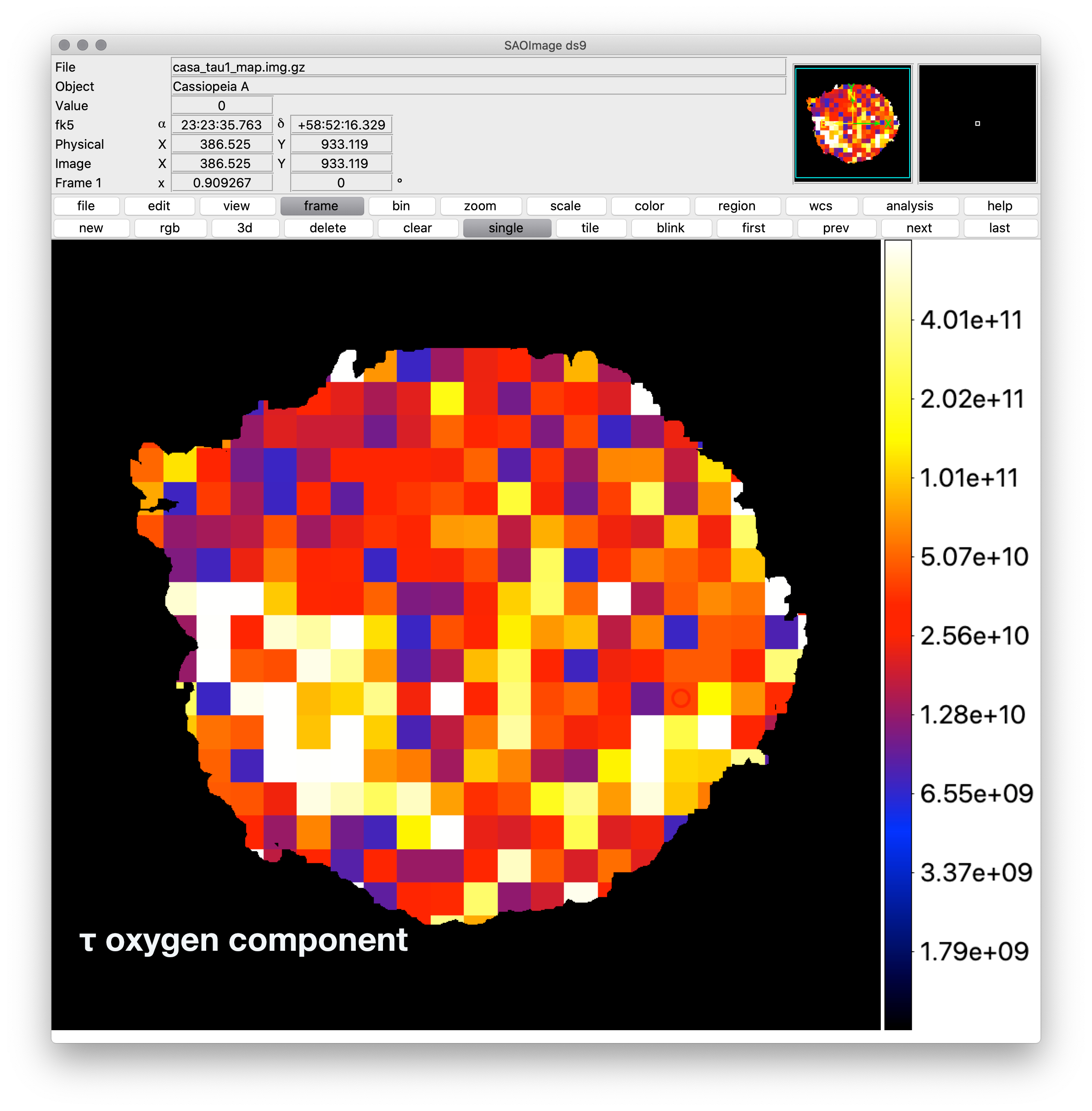}
\includegraphics[trim=55 100 60 263,clip=true,width=0.33\textwidth]{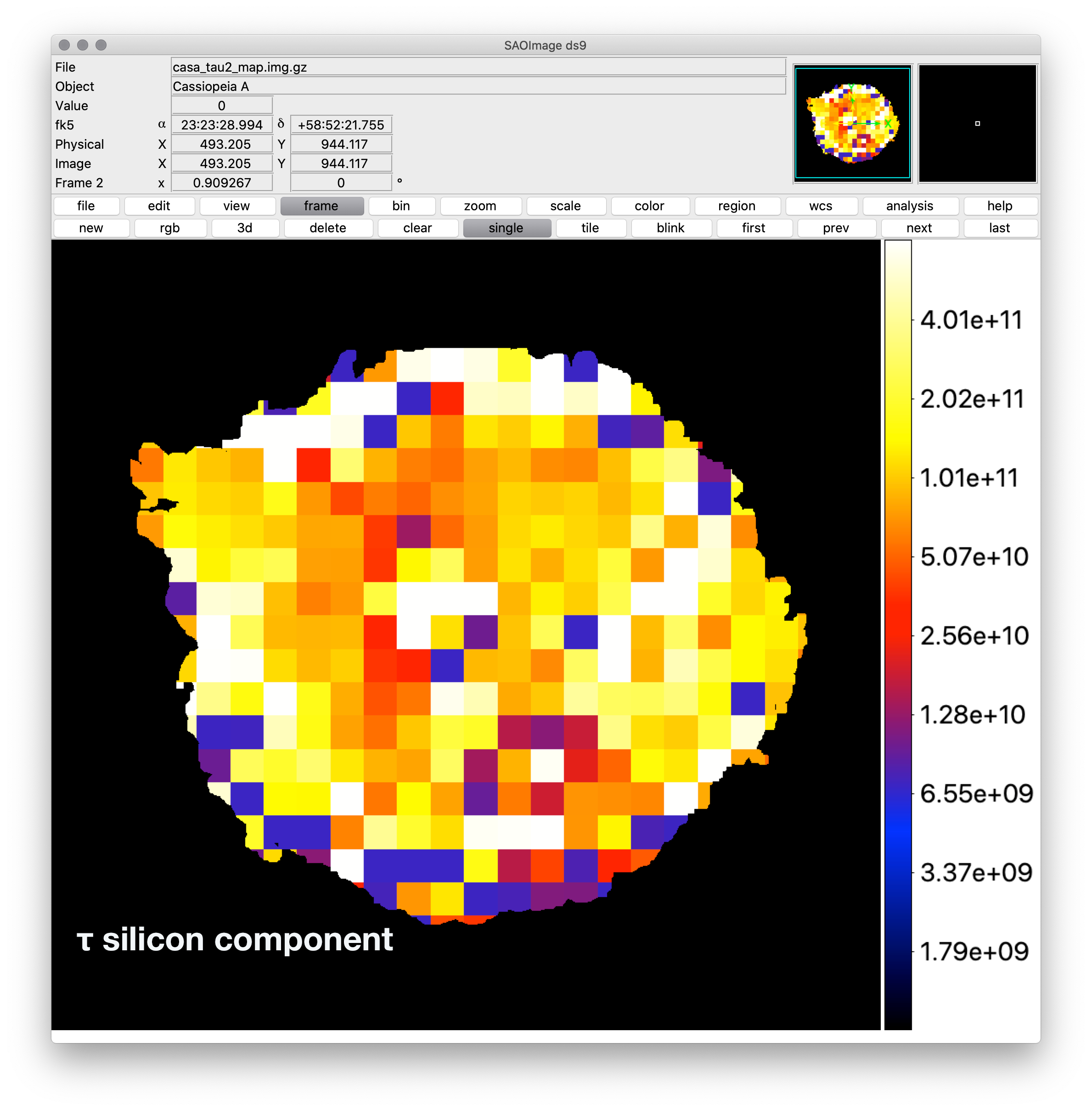}
\includegraphics[trim=55 100 60 263,clip=true,width=0.33\textwidth]{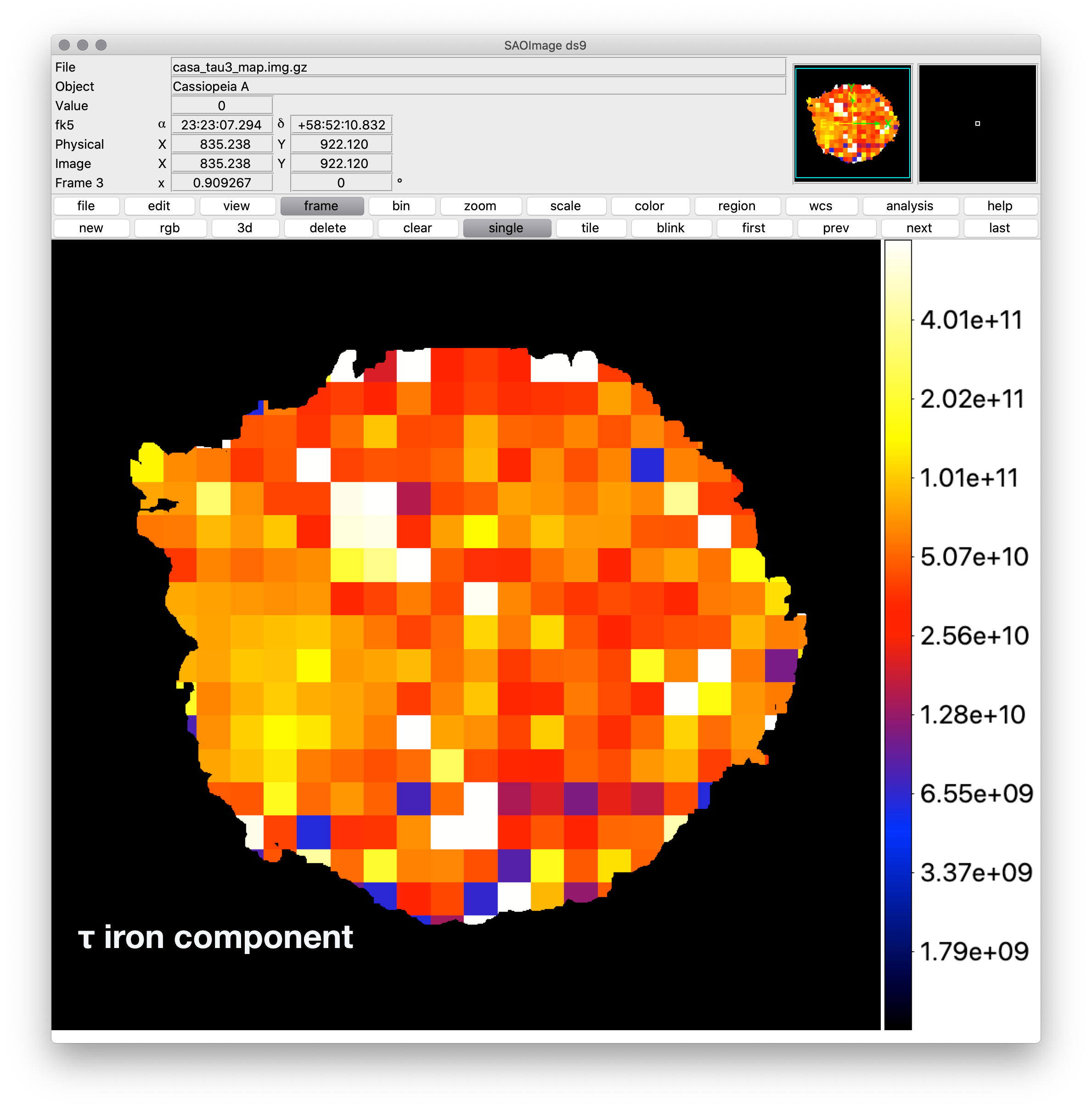}          
          }
            \centerline{
\includegraphics[trim=72 100 60 261,clip=true,width=0.33\textwidth]{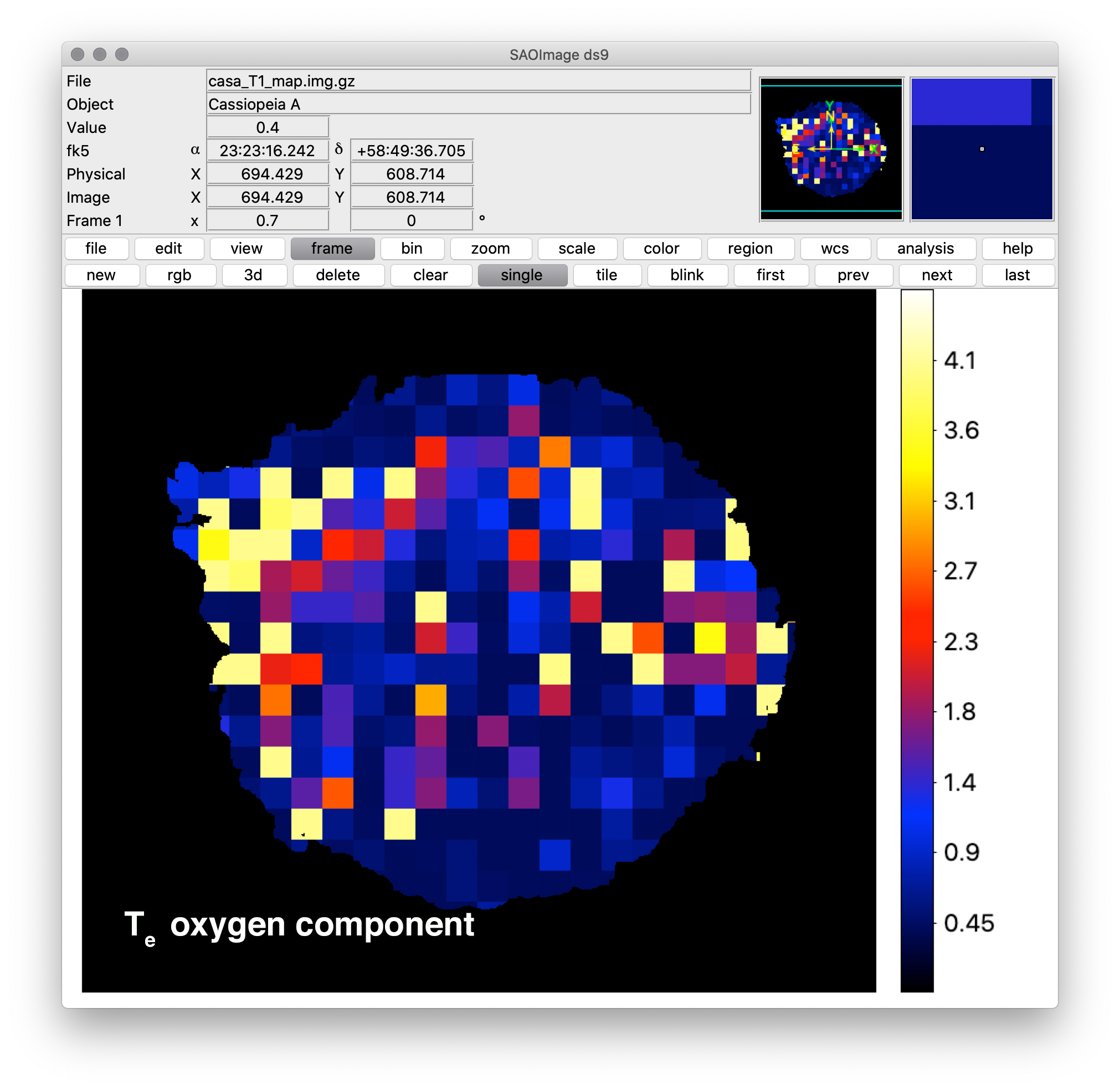}
\includegraphics[trim=72 100 60 261,clip=true,width=0.33\textwidth]{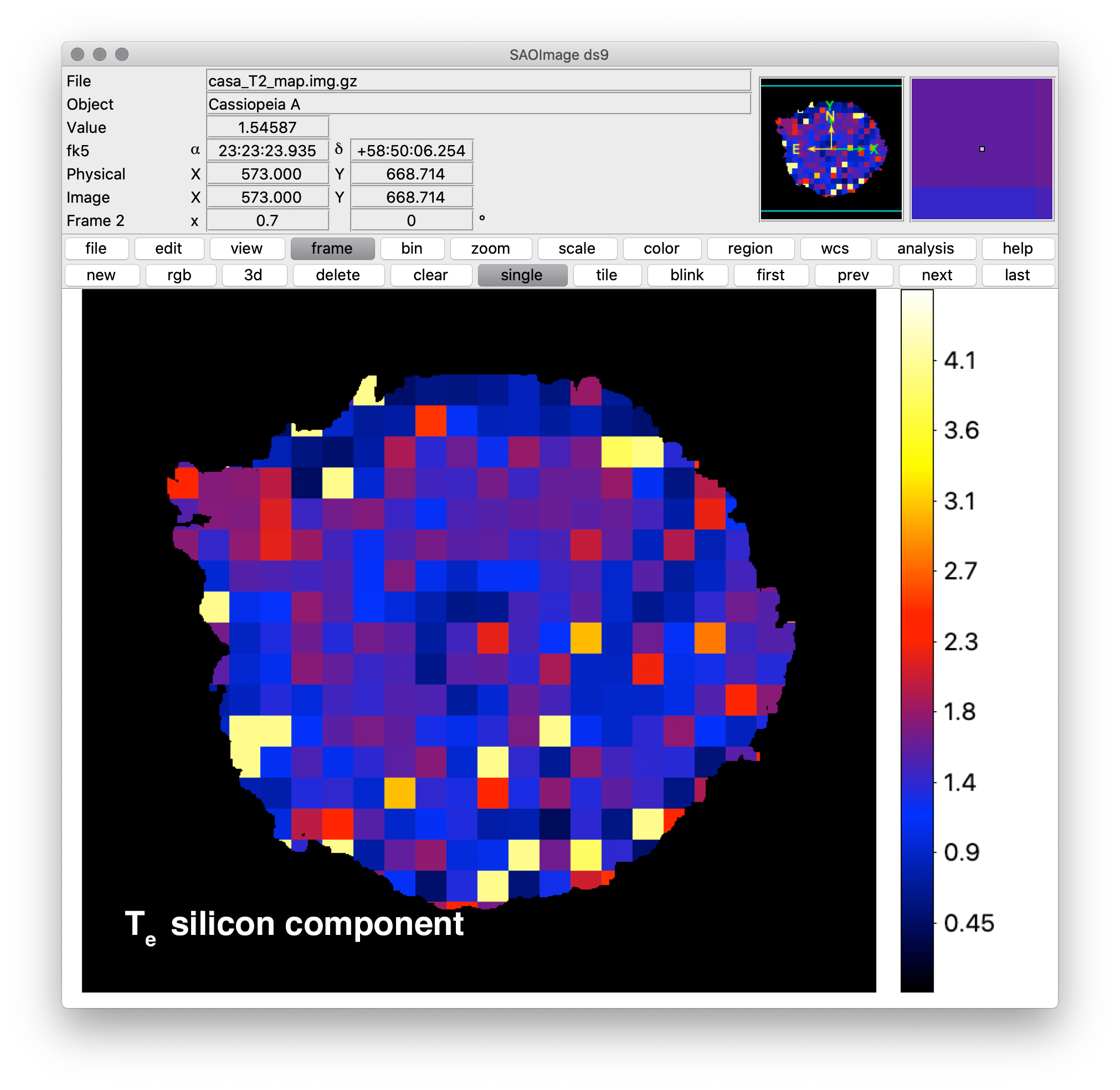}
\includegraphics[trim=72 100 60 261,clip=true,width=0.33\textwidth]{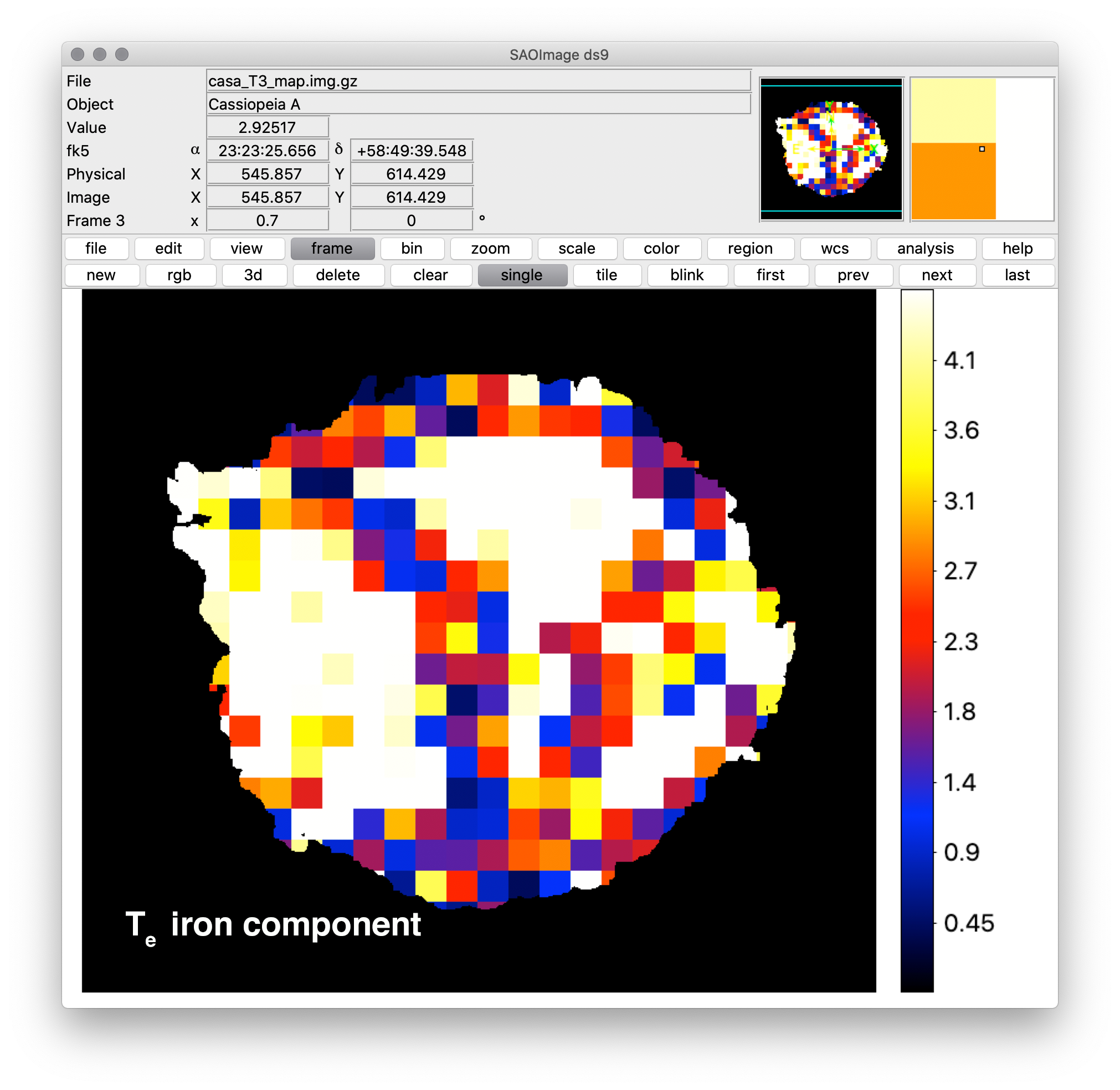}          
          }
           \centerline{
     \includegraphics[trim=72 100 60 263,clip=true,width=0.33\textwidth]{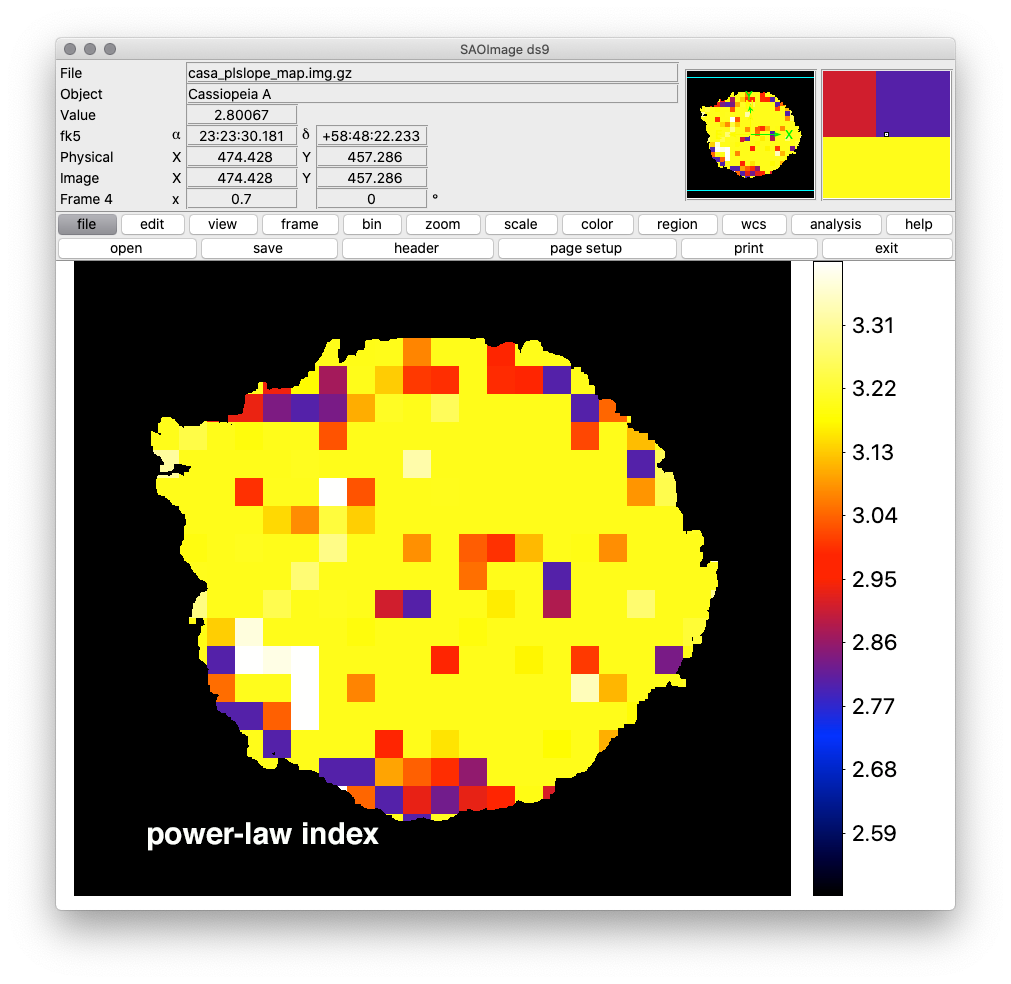}
     \includegraphics[trim=55 100 60 263,clip=true,width=0.33\textwidth]{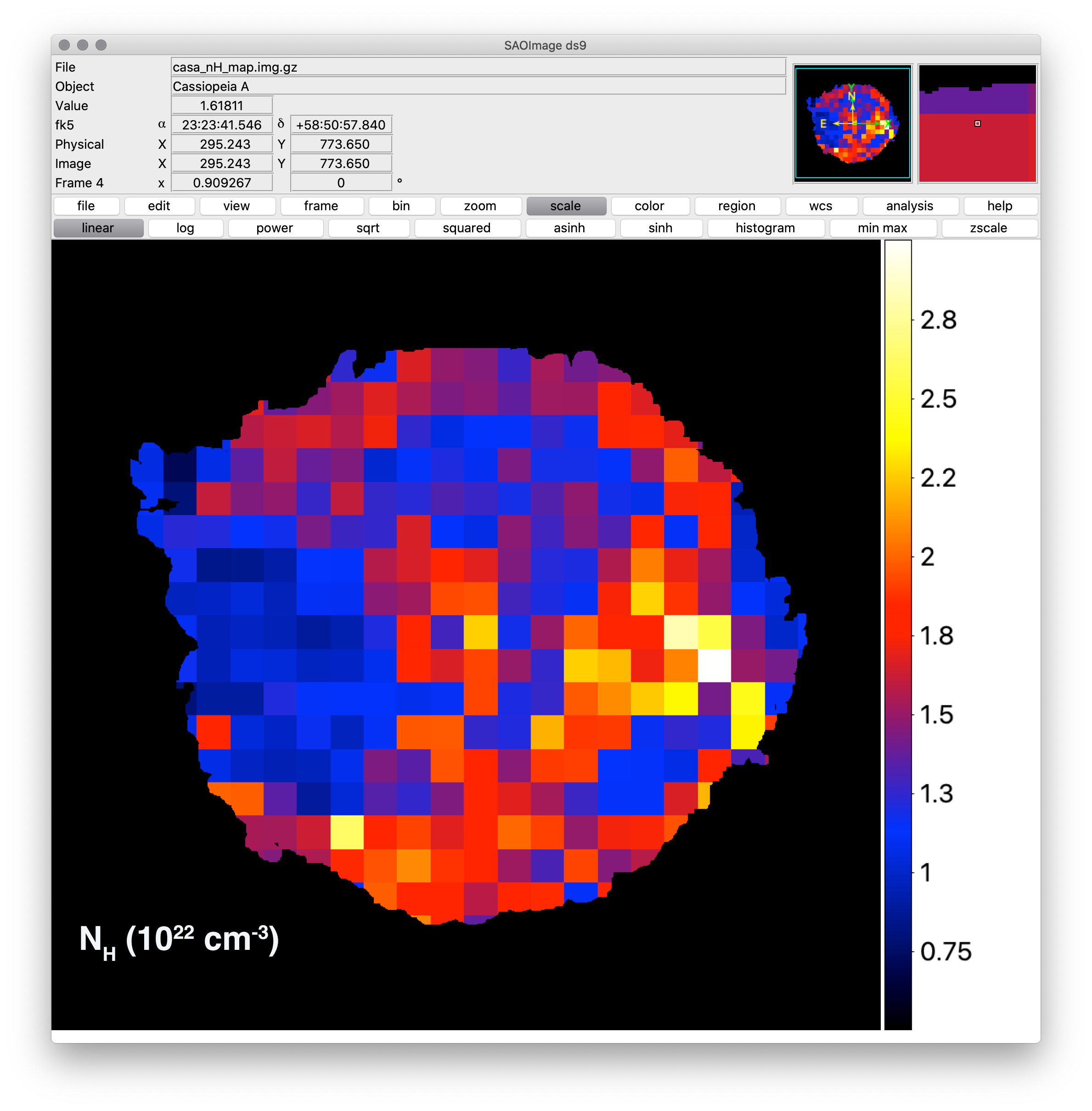}
        \includegraphics[trim=55 100 60 263,clip=true,width=0.33\textwidth]{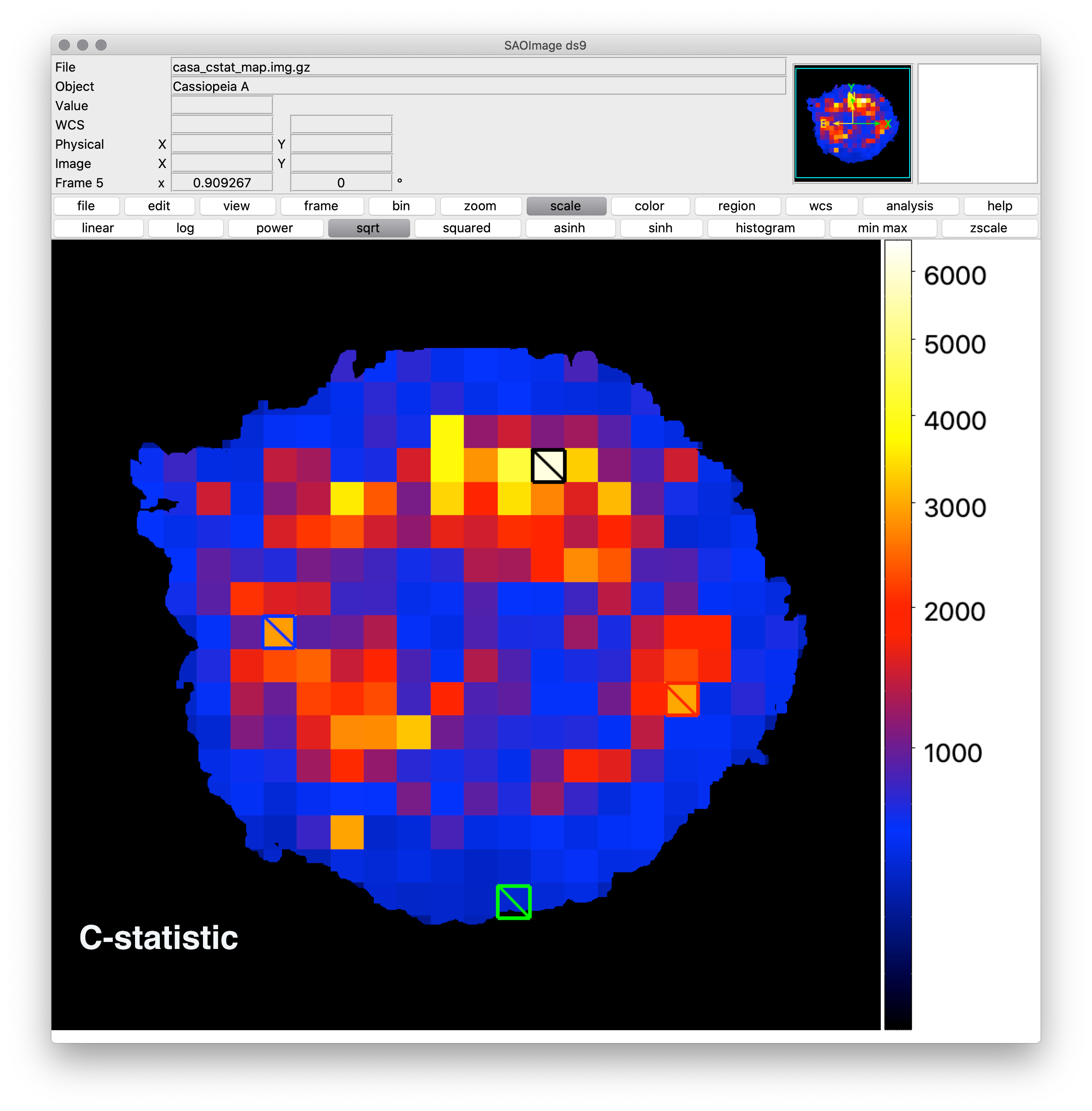}
          }    
          \centerline{
                      \includegraphics[trim=70 80 60 263,clip=true,width=0.33\textwidth]{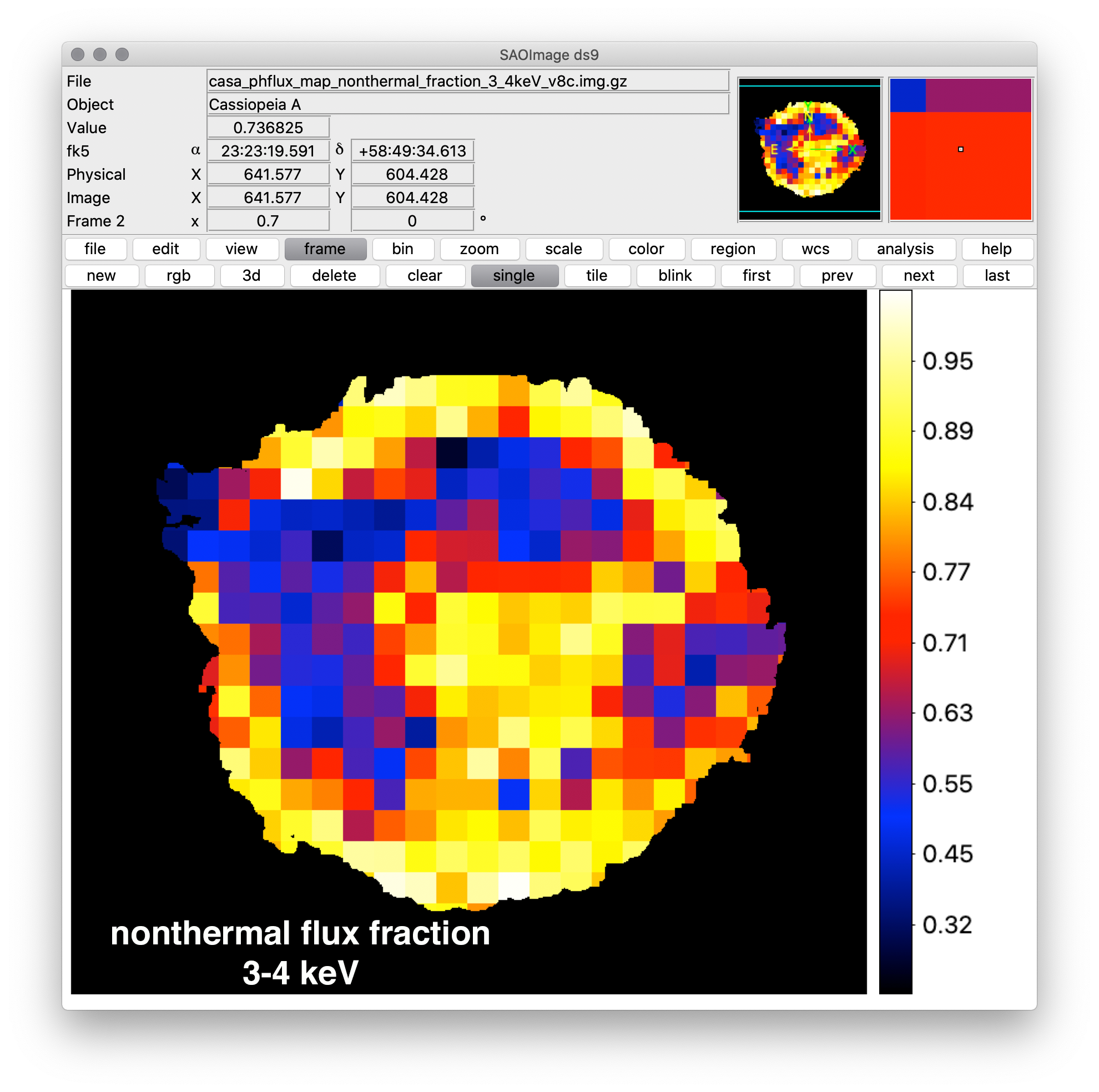}
                      \includegraphics[trim=70 80 60 263,clip=true,width=0.33\textwidth]{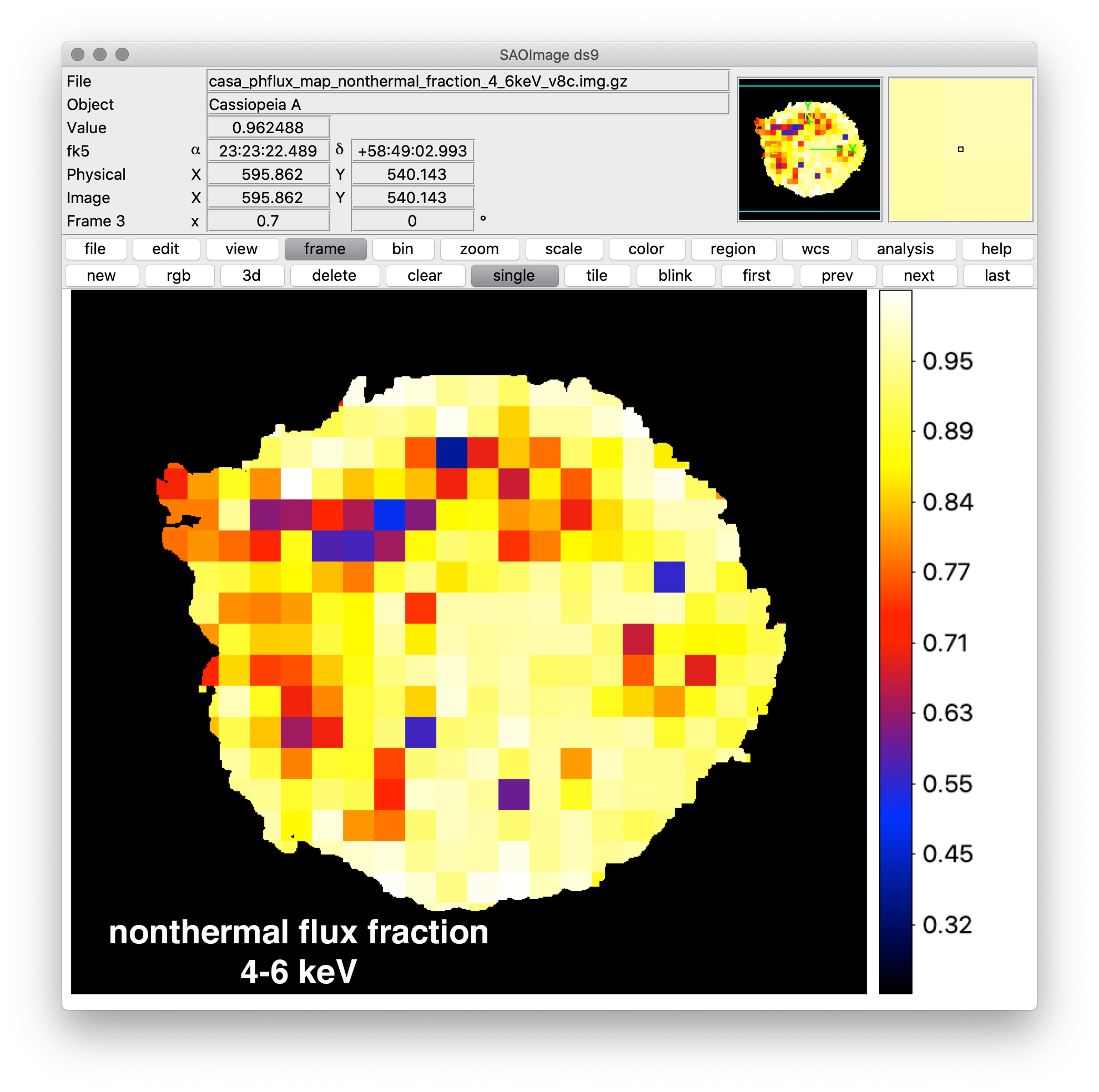}
                      \includegraphics[trim=70 80 60 263,clip=true,width=0.33\textwidth]{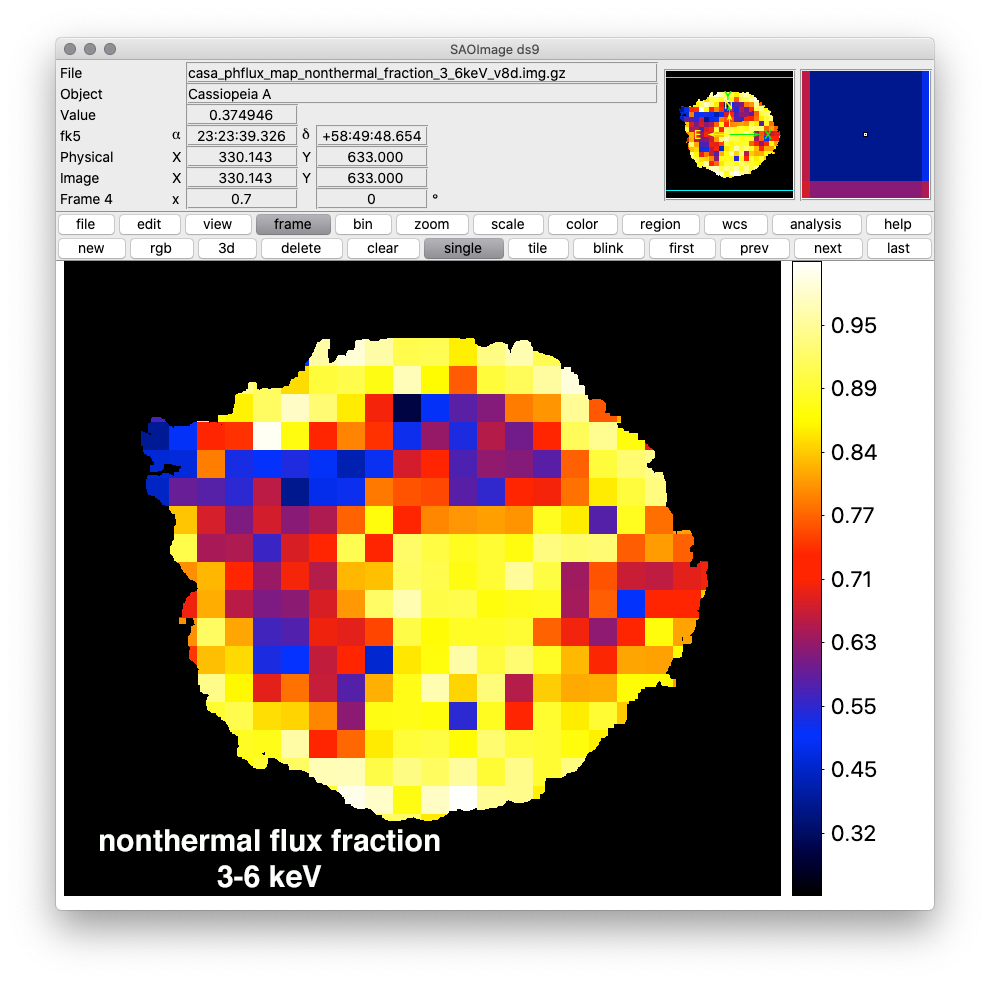}                      
          }
   \caption{
    \label{fig:spectralmaps}
    Examples of the maps of relevant parameters created from the output automated fitting of  Chandra X-ray observations (ObsID 4638).
We used these maps, and similar ones, to generate simulated
    IXPE event lists for assessing polarization degrees of the synchrotron components as a function of position and energy band.
    The top row shows the $\tau=n_{\rm e}t$ values in units of cm$^{-3}$s for the oxygen-rich, silicon-rich (i.e. rich in intermediate mass elements), and iron-rich plasma components, respectively.
    The second row shows the absorption map ($N_{\rm H}$ in cm$^{-2}$), the C-statistic values of the best fits, and the fraction of synchrotron radiation with respect to the overall
    flux in the 3--4 keV,  4--6 keV, and the combined 3--6 keV bands.   
    }
\end{figure}

\begin{figure}
\centerline{         
  \includegraphics[trim=0 0 0 0,clip=true,width=0.75\textwidth]{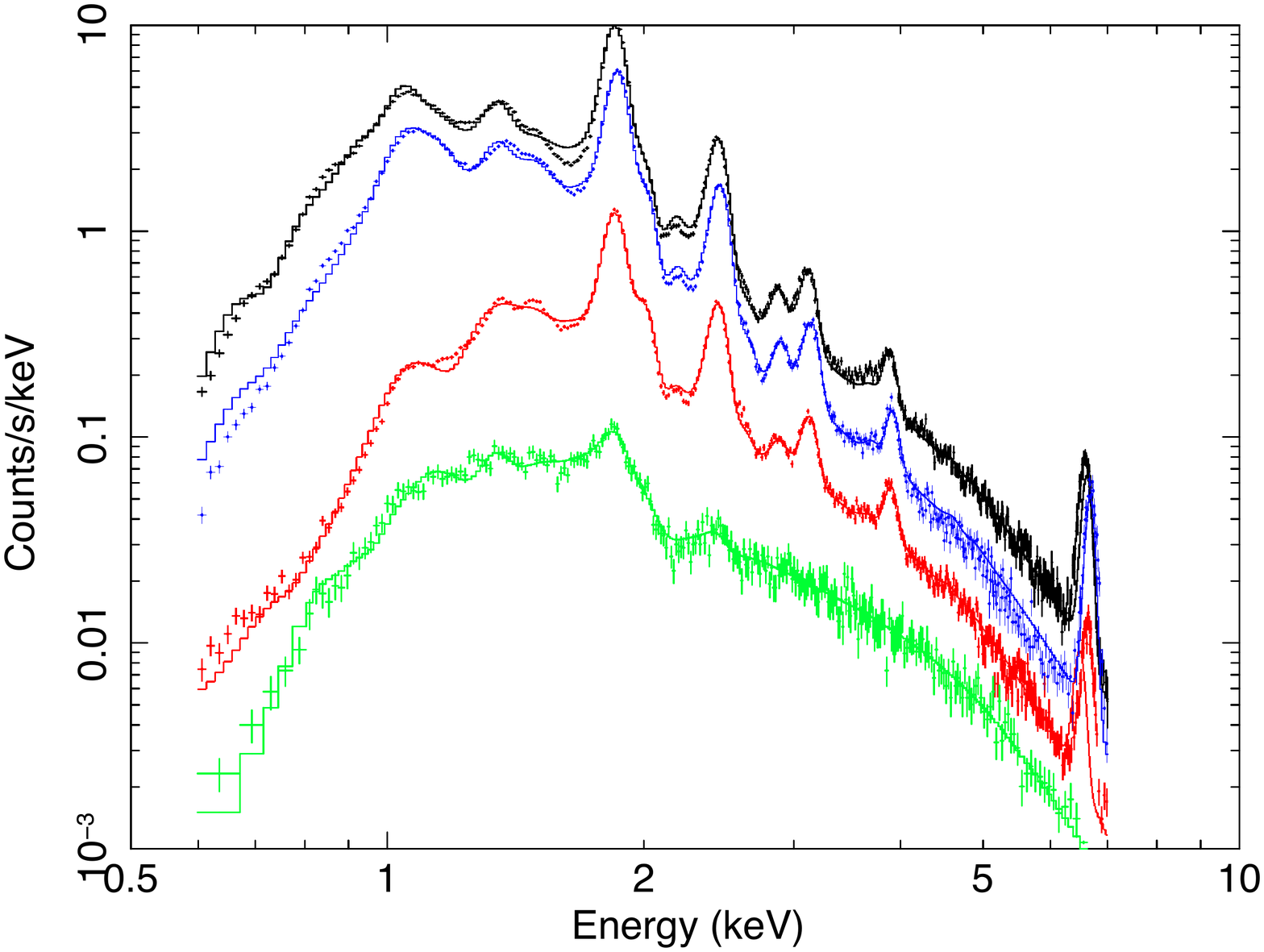}
}
\caption{
\label{fig:examplespec}
Four example spectra and best-fit models based on the automated fitting scripts.
The colors correspond to those of the crossed squares in the C-statistic map in Fig.~\ref{fig:spectralmaps}. 
The black line corresponds to the worst C-statistic. 
The fits are generally satisfactory, but the iron line centroids appear  off in the top three spectra.
}
\end{figure}

\section{Assessing the nonthermal contribution across Cas A}
\label{app:chandra}
The radio emission from Cas A is almost exclusively caused by synchrotron radiation,
but the emission in the 0.5-10 keV consists of  a mixture
of thermal radiation (line emission, bremsstrahlung, free-bound and two-photon emission)
and synchrotron radiation.

At the resolution of the Chandra X-ray observatory, the synchrotron emission
stands morphogically apart \citep{hughes00a,vink03a,hwang04},
being confined to narrow filaments, which may nevertheless
also determine the continuum in projection toward the rest of the SNR.
 The thermal
emission from Cas A is dominated by emission from shocked ejecta 
\citep{laming03}.
In order to prepare for the IXPE observations of Cas A, but also to
estimate the contribution of the synchrotron radiation to each IXPE pixel
as a function of energy, we modeled the emission from Cas A based on fitting
Chandra X-ray spectra from the whole of Cas A,
divided in tiles of 20\arcsec\ by 20\arcsec. We used for the modeling
the {\em xspec} package version 12.12 \citep{arnaud96}, and concentrated on a
single, 164~ks Chandra observation (ObsID 4638) \citep{hwang04}.

The best-fit models per individual tiles were then folded through the spatial
and spectral response functions of IXPE using the \texttt{ixpeobsim} instrumental
response functions. For IXPE simulations we imposed a polarization degree and orientation
on the X-ray synchrotron component. 
Before the launch of IXPE we used these simulations  to test and optimize  the analysis tools used for the results presented in this paper, but at the final
stage, after indeed obtaining detections, we also used the simulations to translate the measured polarization degree 
into a polarization degree for just the synchrotron component. The pre-launch simulations  indicated  that the 3--6~keV band provided
a more sensitive band for detecting polarization than the 4--6~keV (nearly) line-free band, as the broader band considerably improves the statistics.

The model consists
of three thermal, non-equilibrium ionization (NEI), components, each with their own electron temperature for per region, for which we used the {\em vnei} model \citep{borkowski01b},  a power-law
component to fit the synchrotron contribution, and the  Galactic absorption model, {\em tbabs} {\citep[][which was also used for the abundances]{wilms00}}. 
The power-law slope was constrained to be $\Gamma=2.8$--$3.4$, i.e., within $\Delta \Gamma = 0.3$ of
the average spectral index for the synchrotron component of $\Gamma=3.1$ measured by \citet{helder08}.
The three {\em vnei} components represented the metal-dominated ejecta, with the components
representing oxygen-, silicon-, and iron-rich plasma. The abundances of oxygen, silicon and iron elements were set to 10,000 times the solar value
in order to simulate almost pure metal plasmas, which also enhances the contributions of
free-bound and two-photon continuum \citep{greco20}. The oxygen-rich plasma was also responsible for the 
neon- and magnesium-line emission, and the silicon-rich plasma for the intermediate mass elements, but with the
abundances of these elements as  free parameters. Additional free parameters were the temperatures, constrained to 0.4--4 keV (4.5 keV for Fe); 
ionization ages ($n_{\rm e}t$, constrained between $2\times 10^9$--$8\times 10^{11}$~cm$^{-3}$s);  
 the Galactic absorption column ($N_{\rm H}$); the
normalizations of the emission components; and Doppler shifts for the silicon- and iron-rich components.

Maps with the best-fit parameters of these models are shown in
Fig.~\ref{fig:spectralmaps}.\footnote{An of  archive the complete data and model files (xcm files) can be found at \url{https://zenodo.org} (doi:10.5281/zenodo.6597504), which  also includes all data products used for this paper.}
We do not claim that these are best-possible models for the emission
of Cas A, as it is based on an automated procedure, sometimes manually assisted for particularly bad fits. But the final models fit the
Chandra spectra generally well. This can be judged from the example spectral fits in Fig.~\ref{fig:examplespec}, one of which
includes the worst fit in terms of the C-statistic.  Despite the poor statistic, the overall spectral features are captured by the model.
In general we find that the centroid of the Fe-K line emission is not  well fitted. This centroid is determined by the plasma temperature, ionization age and
Doppler shift. However, the ionization age and temperature also determine the iron L-shell emission between 0.8-1.4 keV, and cannot be arbitrarily
varied. So a likely cause of the centroid mismatch is that the Fe-L and Fe-K emission may arise from various plasma components, suggesting that
one plasma component for Fe-rich ejecta is not sufficient.
The maps shows also that the electron temperature of some components tend to get stuck at the predefined upper boundaries,
at which point the fitting algorithm has difficulties further optimizing the fitting results.
This is in particular true
for the electron temperature of the Fe-rich component, for which $T_{\rm e}=4.5$~keV for large parts of the SNR.
The best-fit values for the power-law index are in most cases $\Gamma=3.2$, which is the starting value. This most likely reflects the insensitivity
of the fit to the power-law index under the constrained condition that $\Gamma$ lies within the range 2.8---3.4.
The Galactic absorption map and the ionization age of the Si-rich component compare well
with the ones provided in \citet{hwang12}. However, we note that \citet{hwang12} did not include a full fitting of the nonthermal
component {in} their analysis.

The values from the synchrotron continuum fraction were used to correct the observed polarization degree to the corresponding value for the synchrotron component.
A perhaps surprising by-product of the Chandra spectral mapping is that we find that fits indicate that the contribution of the synchrotron radiation to
the overall continuum emission is surprisingly high, ranging from 38\% to nearly 100\%. \citet{helder08} already showed that this might be the case
based on the extrapolation of the hard X-ray continuum measured by BeppoSAX-PDS to the 4-6~keV band (their Fig.~5), but estimated a more conservative 54\% overall
contribution. Nevertheless, the map of the synchrotron continuum fraction in Fig.~\ref{fig:spectralmaps} matches well morphologically with Fig.~6 (right) in   \citet{helder08} {and the NuSTAR 10--15 keV map in Fig.~6 (right)
in \citet{grefenstette15}.}
For completeness we show in the bottom panels of Fig.~\ref{fig:spectralmaps} the nonthermal flux fraction in the 3--4~keV band (a mixture of line and continuum radiation), the 4--6 keV band (dominated by continuum radiation), and the combined band
3--6~keV band. The latter is the band chosen for the IXPE polarization measurements, based on pre-launch simulations.
Although the 3--4 keV band contains Ar-K and Ca-K line contributions, as can be seen in Fig.~\ref{fig:examplespec}, even in this band the nonthermal continuum fraction is large ($>90$\%) in the X-ray synchrotron
dominated regions, and still 25\%--50\% in the bright shell.

\begin{figure}
\centerline{
\includegraphics[trim=70 80 60 263,clip=true,width=0.35\textwidth]{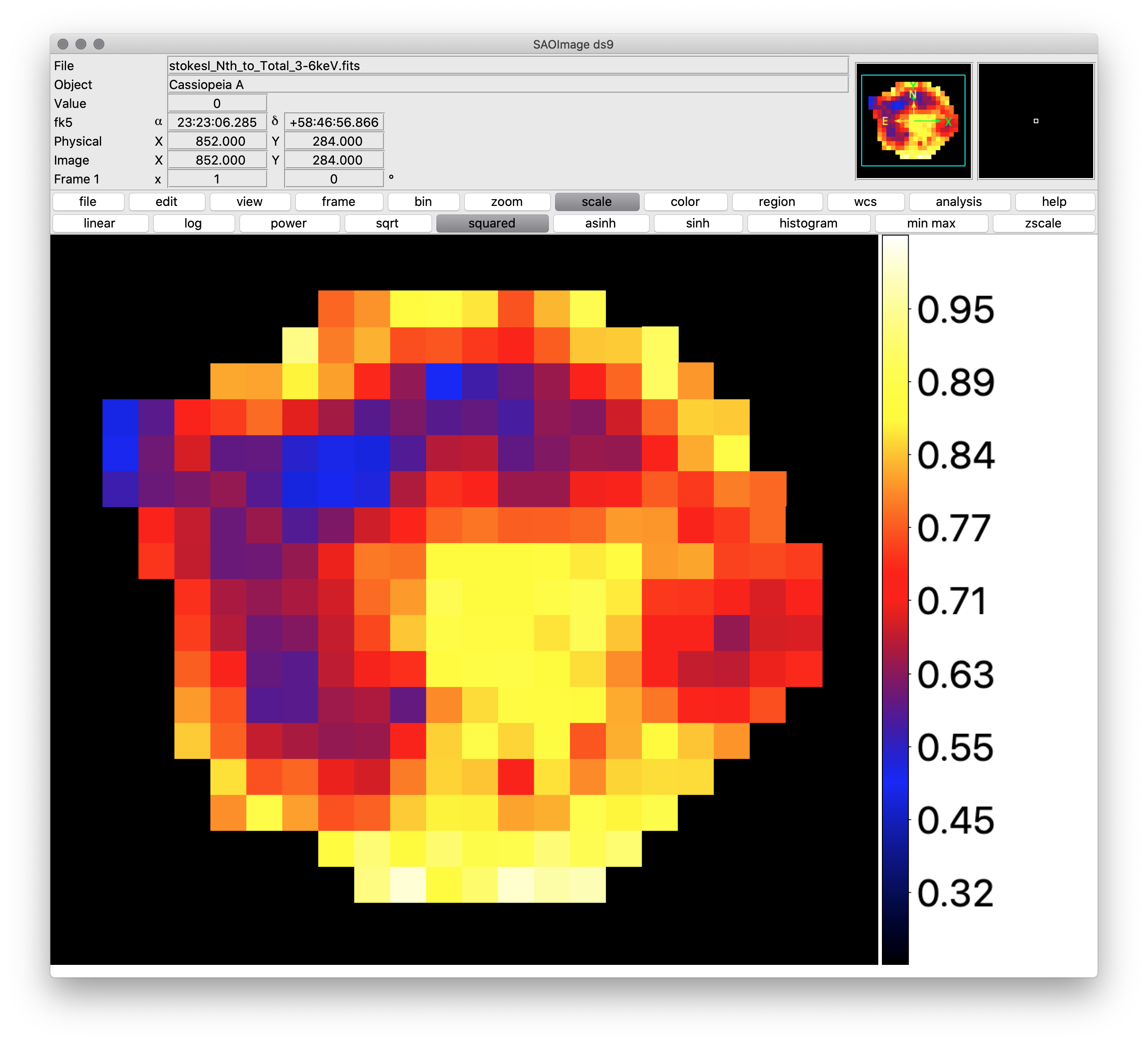}
}
\caption{\label{fig:ixpefraction}
The nonthermal flux fraction in the 3--6 keV band inferred for IXPE, based on data simulated with \texttt{ixpeobsim}. The simulation used the Chandra spectral models as input parameters,  using a full spectral and spatial response modeling---c.f. the last panel
of Fig.~\ref{fig:spectralmaps}.
}
\end{figure}

With \texttt{ixpeobsim} we simulated IXPE event lists for Cas A based on the grid of models for the Chandra spectra. These IXPE simulations showed that the 3--6 keV band is a good compromise between
having a broad enough X-ray band to optimize statistics, and not too much degradation of the polarization signal due to the presence of Ar-K and Ca-K lines, but avoiding the very bright Si-K and S-K lines.
Some Si-K and S-K line emission may be leaking into the 3--6 keV band due to the low spectral resolution of IXPE. The nonthermal flux fraction  in the 3--6 keV inferred for IXPE is displayed in Fig.~\ref{fig:ixpefraction},
which includes all instrumental effects.

\section{Polarization signal as a function of chosen energy band}
\label{app:bandeffects}

The main text concentrates on the IXPE polarization signal analysis in the 3--6 keV band. In Table~\ref{tab:polbreakdown} we show  for all regions combined and the FS+ RSW region the break-down
in polarization signal for the 3--4 keV and 4--6 keV band, as well as the inferred correction factor needed to obtain the polarization signal of the synchrotron component only, as based
on the models in Appendix~\ref{app:chandra}. All values refer to measurements based on the  Stokes Q$^\prime$ and U$^\prime$ parameters, for which the definition of the polarization angle changes
as a function of pixel position, assuming a circular symmetry with respect to the center of the SNR
---see section~\ref{sec:aligned} for details.

Note that for the continuum-dominated, 4--6~keV band the polarization detection significance is 4.1$\sigma$ for all regions combined and 4.6$\sigma$ for the FS+RWS region.
Although at a  lower significance, the measurements in the continuum-dominated band supports the overall low polarization degree reported in the main text.

\begin{table}
\caption{Polarization signal as a function of selected energy band.
\label{tab:polbreakdown}
}
\begin{tabular}{lcccccc}\hline\hline\noalign{\smallskip}
Band    & MPD99 & syn. corr.\tablenotemark{a} &  Pol. Degree& Angle & Signficance \\
 (keV)   & (\%)       &                                         & (\%)  & ($^\circ$) &\\ \noalign{\smallskip}\hline
 \multicolumn{6}{c}{All regions combined}\\
 3--4 &1.1 & 1.5 & $1.2\pm 0.4$ & $85.1\pm 8.8$ & 2.6$\sigma$\\ 
 4--6 &  1.8 & 1.2 & $2.9\pm  0.6$ & $86.1\pm 6.1$  & 4.1$\sigma$\\
 3--6 & 1.0 & 1.4 & $1.8\pm 0.3$  & $85.7\pm5.3$ & 4.9$\sigma$\\
  \multicolumn{6}{c}{FS+RWS}\\
   3--4 &1.9 & 1.4 & $1.8\pm 0.6$ & $78.0\pm 10.3$ & 2.2$\sigma$\\ 
 4--6 &  3.0 & 1.2 & $5.0\pm  1.0$ & $92.1\pm 5.6$  & 4.6$\sigma$\\
 3--6 & 1.7 & 1.3 & $3.0\pm 0.6$  & $87.2\pm5.4$ & 4.8$\sigma$\\
 \noalign{\smallskip}\hline
\end{tabular}
\tablenotetext{a}{Synchrotron correction fraction, i.e. the ratio of the inferred total flux  over the nonthermal flux as based on the Chandra model fits.}
\end{table}

\end{document}